\documentclass[
prd %, twocolumn%
,showpacs ,amssymb,superscriptaddress,aps
]{revtex4}
\usepackage{graphicx}
\input{epsf}

\usepackage{amsmath,amssymb}
\usepackage{bm}
\usepackage{times}

\newcommand{\dalm}{\kern1pt\vbox{\hrule height 0.9pt\hbox{\vrule width 0.9pt
\hskip 2.5pt\vbox{\vskip 5.5pt}\hskip 3pt\vrule width 0.3pt}\hrule height 0.3pt}
\kern1pt}

\begin{document}

%\twocolumn[\hsize\textwidth\columnwidth\hsize\csname @twocolumnfals\endcsname

% For two column
%\wideabs{

\title{Magnetized relativistic stellar models in Eddington-inspired Born-Infeld gravity}

\author{Hajime Sotani}
\email{sotani@yukawa.kyoto-u.ac.jp}
%\affiliation{Yukawa Institute for Theoretical Physics, Kyoto University, Kyoto 606-8502, Japan}
\affiliation{Division of Theoretical Astronomy, National Astronomical Observatory of Japan, 2-21-1 Osawa, Mitaka, Tokyo 181-8588, Japan}

\date{\today}

% Abstract
\begin{abstract}
We consider the structure of the magnetic fields inside the neutron stars in Eddington-inspired Born-Infeld (EiBI) gravity. In order to construct the magnetic fields, we derive the relativistic Grad-Shafranov equation in EiBI, and numerically determine the magnetic distribution in such a way that the interior magnetic fields should be connected to the exterior distribution. Then, we find that the magnetic distribution inside the neutron stars in EiBI is qualitatively similar to that in general relativity, where the deviation of magnetic distribution in EiBI from that in general relativity is almost comparable to uncertainty due to the equation of state (EOS) for the neutron star matter. 
However, we also find that the magnetic fields in the crust region are almost independent of the coupling constant in EiBI, which suggests a possibility to obtain the information about the crust EOS independently of the gravitational theory via the observations of the phenomena associated with the crust region. In any case, since the imprint of EiBI gravity on the magnetic fields is weak, the magnetic fields could be a poor probe of gravitational theories, considering many magnetic uncertainties.
%However, we also find that the magnetic fields in the crust region for the neutron star with canonical mass depend weakly on the coupling constant in EiBI, while some magnetic properties in the low-mass neutron star depend weakly on the EOS for neutron star matter. Thus, via the observations of the phenomena associated with the crust region and/or low-mass neutron stars, it might be possible not only to obtain the information about the EOS for neutron star matter, but also to probe the gravitational theory. 
\end{abstract}

\pacs{04.40.Dg, 04.50.Kd, 04.40.Nr}
%
%%%%%%%%%%%%%%%%%%%%%%%%%%%%%%%%%%%%%%%%%%%%%%%%%
%  04.40.Dg :  Relativistic stars: structure, stability, and oscillations (see also 97.60.-s Late stages of stellar evolution) 
%  04.50.Kd  : Modified theories of gravity
%  04.40.Nr  :  Einstein-Maxwell spacetimes, spacetimes with fluids, radiation or classical fields
%  04.80.Cc  : Experimental tests of gravitational theories
%%%%%%%%%%%%%%%%%%%%%%%%%%%%%%%%%%%%%%%%%%%%%%%%%
%]
% For two column
%}
\maketitle
%\baselineskip 24pt
%%%%%%%%%%%%%%%%%%%%%%%%%%%%%%%%%%%%%%%%%%%%%%%%
\section{Introduction}
\label{sec:I}
%%%%%%%%%%%%%%%%%%%%%%%%%%%%%%%%%%%%%%%%%%%%%%%%

The magnetic field is one of the principal properties in the phenomena of the astrophysical objects. In fact, it is believed that the magnetic fields can play an important role during supernova explosions, gamma-ray bursts, jets from active galactic nuclei, and so on.  The existence of strongly magnetized neutron stars, the so-called magnetars, is also suggested via the measurements of spin period and its down rate of the central objects in soft gamma repeaters (SGRs) and anomalous X-ray pulsars. According to the magnetic dipole model, the strength of surface magnetic fields of the magnetars is considered to be as large as $10^{14}-10^{15}$ G \cite{K1998,H1999}. From the SGRs, sporadic radiations of $\gamma$- and X-rays are observed, while fierce flare activities called the giant flares are also detected on rare occasions. In particular, the quasi-periodic oscillations (QPOs) discovered in the afterglow of the giant flares give us the evidences of the oscillations of the magnetized neutron stars \cite{SW2006}. To theoretically explain the QPO frequencies, there are many attempts in term of the crustal oscillations \cite{SW2009,S2011a,GNHL2011,SNIO2012,SNIO2013} and/or magnetic oscillations \cite{SKS2007,SK2009,CK2011,GCNFM2012,PL2014}. In any case, in addition to the EOS for neutron star matter, the structure of magnetic fields inside the neutron star must be crucial to understand such phenomena.

On the other hand, the gravitational theory must be also imperative to discuss the relativistic objects. The general relativity is mathematically beautiful theory of gravity, and its validity has been probed by a lot of experiments and astronomical observations. However, most of the verifications of general relativity have been done in a weak field regime, such as the Solar System \cite{W1993}, while the tests in a strong field regime are very poor. Perhaps, the gravitational theory describing the astronomical phenomena in a strong field regime might be different from general relativity. This is a reason why modified theories of gravity are proposed. Since the observable properties could depend on the gravitational theory, one would see the imprint of the gravitational theory as an inverse problem \cite{B2015}. In fact, the science technology is developing increasingly, which will enable us to observe the relativistic objects and phenomena around such objects with high precision. Probably, the gravitational waves radiated from such a system are also one of them. Through these observations, it is possible to probe the gravitational theory \cite{P2008,SK2004,S2009}.

As a modified theory of gravity, EiBI is recently drawing attention in the context of the avoidance of the big bang singularity \cite{AF2012,BFS12013}. This theory is originally proposed by Ba\~nados and Ferreira \cite{EiBI}, based on the gravitational action proposed by Eddington \cite{E1924} and on the nonlinear electrodynamics by Born and Infeld \cite{BI}. EiBI is developed according to a Paratini approach, where the connection is considered as an independent field, because the field equations contain ghosts in the metric approach \cite{DG1998}. The deviation of EiBI from general relativity can be seen only when the matter exists, i.e., EiBI in vacuum is completely equivalent to general relativity, and the deviation becomes significant in high density region. Thus, the compact objects are good candidates to see such a deviation. Up to now, the compact objects in EiBI are discussed on several occasions and shown the deviation in stellar properties from the expectations in general relativity \cite{PCD2011,PDC2012,SLL2012,SLL2013,HLMS2013,S2014a,S2014b}. Maybe, via the direct observations of such stellar properties, one would distinguish EiBI from general relativity. We remark that in EiBI the curvature singularity can appear at the stellar surface for polytropic EOSs \cite{PS2012}, which must be a problem to solve even thought this theory is attractive.

However, in spite of the importance of magnetic effects in astronomical phenomena, the magnetic fields on the neutron stars in EiBI have not been considered. There are solely the considerations on the electrically charged black holes in EiBI \cite{EiBI,Wei2014,SM2014}. Thus, in this paper, we consider the magnetized relativistic stellar models in EiBI. This discussion could become a fist step to examine the phenomena associated with the neutron stars in EiBI. Actually, there are many uncertainties in the magnetic properties, such as its geometry and the currents supporting it, even for a given fixed EOS. So, it must be quite difficult to see the imprints of the gravitational theory in the magnetic properties, if neutron stars would have different magnetic geometry and/or crust properties irrespectively of the theory of gravity. In this paper, to see how the magnetic fields depend on the gravitational theory, we especially focus on
%to avoid such diversity in interior properties, we assume that the magnetic geometry and the interior properties of any neutron stars would eventually approach a specific model, which should still depend on the gravitational theory. On this assumption, we might have a chance to see the imprint of the gravitational theory observationally. In particular, we focus on 
the axisymmetric dipole configuration of magnetic fields, because such configuration could be dominant in the old neutron stars.
The additional factors to determine the magnetic properties should be taken into account, but we neglect such effects here to simplify the problem. In this paper, we adopt geometric units, $c=G=1$, where $c$ and $G$ denote the speed of light and the gravitational constant, respectively, and the metric signature is $(-,+,+,+)$.

%%%%%%%%%%%%%%%%%%%%%%%%%%%%%%%%%%%%%%%%%%%%%%%%
\section{Magnetized Stellar models in EiBI}
\label{sec:II}
%%%%%%%%%%%%%%%%%%%%%%%%%%%%%%%%%%%%%%%%%%%%%%%%

Before considering the stellar models in EiBI, we briery mention EiBI. This gravitational theory is obtained from the action $S$ given by
\begin{equation}
  S=\frac{1}{16\pi}\frac{2}{\kappa}\int d^4x \left(\sqrt{|g_{\mu\nu} + \kappa R_{\mu\nu}|} - \lambda\sqrt{-g}\right) + S_{\rm M}[g,\Psi_{\rm M}],
\end{equation}
where $g$ and $|g_{\mu\nu} + \kappa R_{\mu\nu}|$ denote the determinants of the physical metric $g_{\mu\nu}$ and $(g_{\mu\nu} + \kappa R_{\mu\nu})$, $R_{\mu\nu}$ is the Ricci tensor constructed from the connection $\Gamma^\mu_{\alpha\beta}$, and $S_{\rm M}$ denotes the matter action depending on the metric $g_{\mu\nu}$ and matter field $\Psi_{\rm M}$. 
That is, the matter field is assumed to minimally couple to the metric tensor, $g_{\mu\nu}$, i.e., the matter action depends on $g_{\mu\nu}$ independently of the connection $\Gamma$.
This theory also has a dimensionless constant $\lambda$ and the Eddington parameter $\kappa$, which are related to the cosmological constant as $\Lambda=(\lambda - 1)/\kappa$. Since we especially focus on the asymptotic flat solutions ($\Lambda=0$) in this paper, hereafter we take $\lambda = 1$. On the other hand, $\kappa$ is constrained from the observations in solar system, big bang nucleosynthesis, and the existence of neutron stars \cite{EiBI,PCD2011,kappa01,kappa02}. The existence of neutron stars can give us the strong constraint on $\kappa$, i.e., $|\kappa| \lesssim 1$ m$^5$ kg$^{-1}$ s$^{-2}$ \cite{PCD2011}. Recently, the possibility to constrain $\kappa$ with the terrestrial measurements of the neutron skin thickness of ${}^{208}$Pb and the astronomical observations of the radius of $0.5M_\odot$ neutron star, is also suggested \cite{S2014a}.  In this paper, we adopt the normalized coupling constant such as $8\pi\kappa\varepsilon_s$, where $\varepsilon_s$ denotes the saturation density, i.e., $\varepsilon_s=2.68\times 10^{14}$ g/cm$^3$. We remark that $8\pi\kappa\varepsilon_s$ becomes a dimensionless parameter.

EiBI is characterized by two independent fields, i.e., the physical metric $g_{\mu\nu}$ and the connection $\Gamma^\mu_{\alpha\beta}$. So, varying the action with respect to $\Gamma^{\mu}_{\alpha\beta}$ and $g_{\mu\nu}$, one can obtain the field equations for $\lambda=1$;
\begin{gather}
  q_{\mu\nu} = g_{\mu\nu} + \kappa R_{\mu\nu}, \label{eq:2} \\
  \sqrt{-q}q^{\mu\nu} = \sqrt{-g}g^{\mu\nu} - 8\pi\kappa\sqrt{-g}T^{\mu\nu}, \label{eq:3}
\end{gather}
where $q$ is determinant of $q_{\mu\nu}$ and $q_{\mu\nu}$ is an auxiliary metric associated with the connection as $\Gamma^\mu_{\alpha\beta} = q^{\mu\sigma}\left(q_{\sigma\alpha,\beta} + q_{\sigma\beta,\alpha} - q_{\alpha\beta,\sigma}\right)/2$. $T^{\mu\nu}$ denotes the energy-momentum tensor, which is given by $T^{\mu\nu}=(\delta S_{\rm M}/\delta g_{\mu\nu})/\sqrt{-g}$. Equation (\ref{eq:3}) shows that the auxiliary metric $q_{\mu\nu}$ becomes equivalent to the physical metric $g_{\mu\nu}$, if $T^{\mu\nu}=0$. That is, EiBI without matter reduces to general relativity in vacuum \cite{EiBI}. In addition to the above field equations, the energy-momentum tensor should satisfy the conservation law, i.e., $\nabla_\mu T^{\mu\nu}=0$, where the covariant derivative $\nabla_\mu$ is defined by $g_{\mu\nu}$.
%We remark that the energy conservation law can be reduced from the field equation (\ref{eq:3}) \cite{EiBI}. Since Eq. (\ref{eq:3}) is the equation of motion derived by varying with respect to $g_{\mu\nu}$, one can consider that the metric $g_{\mu\nu}$ not $q_{\mu\nu}$ as the physical metric.
%because the matter field is minimally coupled to the metric $g_{\mu\nu}$ \cite{WGR}.
As far as we know, unfortunately, there is no explicit proof that the conservation law of $\nabla_\mu T^{\mu\nu}=0$ is directly derived from the field equations (\ref{eq:2}) and (\ref{eq:3}). However, since the matter field is minimally coupled to the metric $g_{\mu\nu}$, the conservation law might be obtained as in Ref. \cite{WGR}, if the argument in \cite{WGR} is applicable even for a bi-metric theory like EiBI.

\begin{figure*}
\begin{center}
\begin{tabular}{cc}
\includegraphics[scale=0.5]{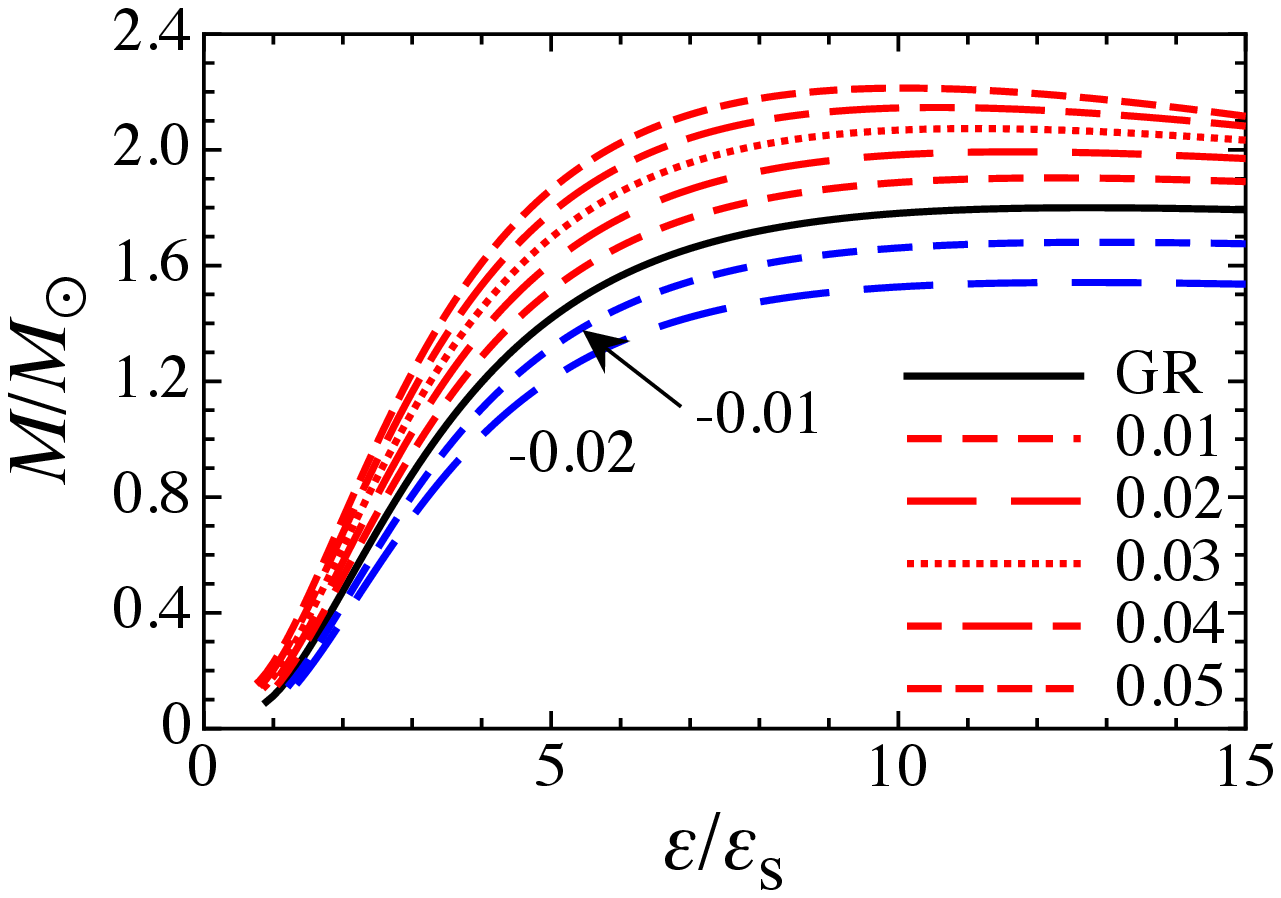} &
\includegraphics[scale=0.5]{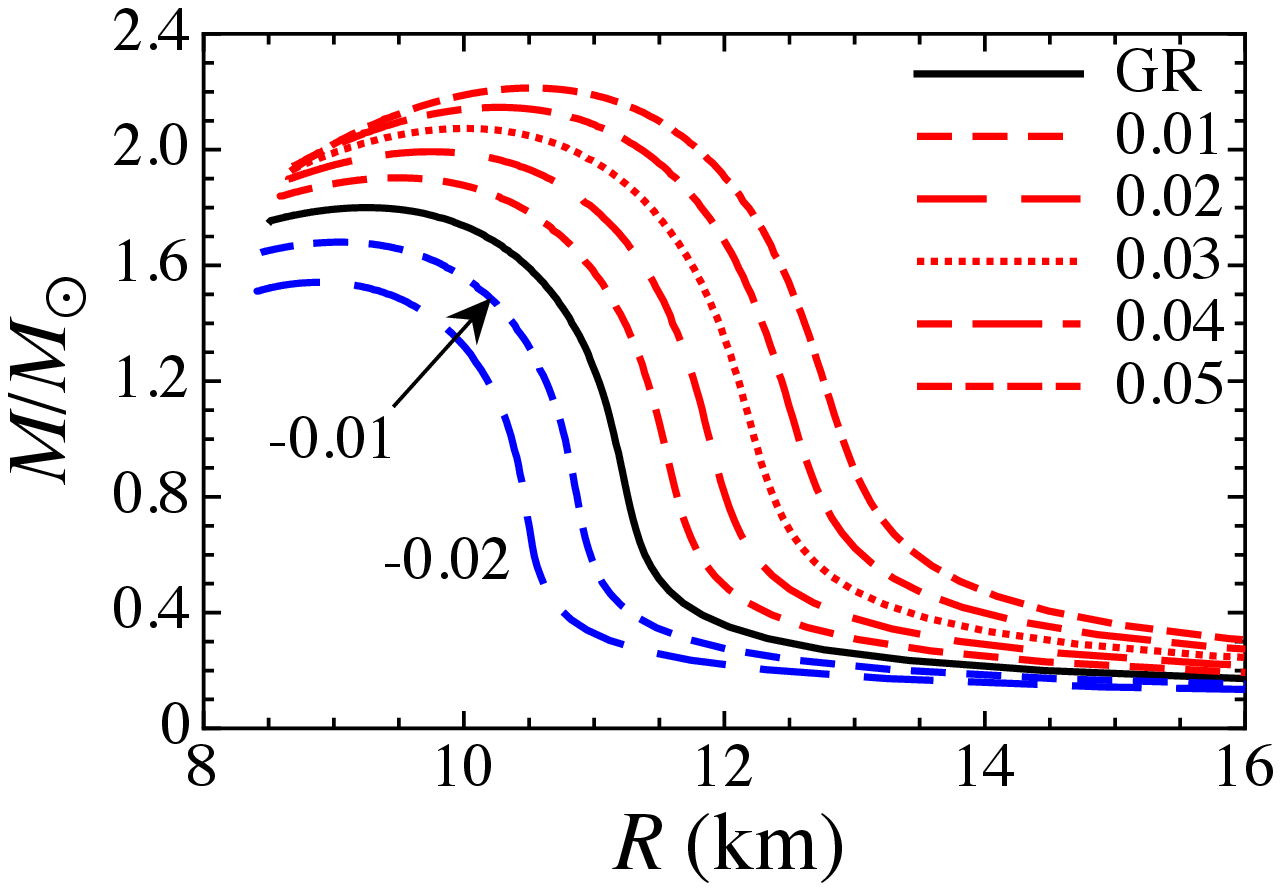}
\end{tabular}
\end{center}
\caption{%%
Neutron star models in EiBI with FPS EOS. The left panel corresponds to the stellar mass as a function of the central density normalized by the saturation density, while the right panel corresponds to the stellar mass as a function of the stellar radius. The solid line denotes the result in general relativity ($\kappa=0$) and the other lines denote the results in EiBI with various normalized coupling constant $8\pi\kappa\varepsilon_s$.
}
\label{fig:MR}
\end{figure*}

Now, we consider the neutron star models in EiBI. In general, magnetized neutron stars could deform due to non-spherically symmetric magnetic pressure. However, the magnetic energy in the neutron star is much smaller than the gravitational binding energy even for a magnetar, which is strongly magnetized neutron star. That is, the deformation due to the magnetic pressure is quite small and the shape of star is almost spherically symmetric. Thus, in this paper, we neglect the stellar deformation induced by the existence of magnetic field. Under such assumption, the equilibrium stellar model can be determined as a solution of Tolman-Oppenheimer-Volkov (TOV) equations in EiBI \cite{PCD2011,PDC2012,SLL2012,SLL2013,HLMS2013,S2014a}. The metric describing the stellar models is given by
\begin{gather}
  g_{\mu\nu}dx^\mu dx^\nu = -e^{\nu(r)}dt^2 + e^{\lambda(r)}dr^2 + f(r)d\Omega^2, \\
  q_{\mu\nu}dx^\mu dx^\nu = -e^{\beta(r)}dt^2 + e^{\alpha(r)}dr^2 + r^2d\Omega^2,
\end{gather}
where $d\Omega^2=d\theta^2 + \sin^2\theta d\phi^2$. In this paper, we consider the stellar models composed of perfect fluid, i.e., $T^{\mu\nu}=(\varepsilon + p)u^\mu u^\nu + pg^{\mu\nu}$, where $\varepsilon$, $p$, and $u^\mu$ are the energy density, pressure, and four velocity of matter given by $u^\mu=(e^{-\nu/2},0,0,0)$. Then, one can show that $abf=r^2$ from Eq. (\ref{eq:3}), where $a$ and $b$ are given by $a=\sqrt{1+8\pi\kappa\varepsilon}$ and $b=\sqrt{1-8\pi\kappa p}$, respectively \cite{S2014a}. In addition to TOV equations, one needs to prepare the EOS for neutron star matter to construct the stellar models. We particularly adopt FPS \cite{FPS} and SLy4 EOSs \cite{SLy4}, which are  based on the Skyrme-type effective interaction (also see \cite{SIOO2014} for the adopted EOSs). Figure \ref{fig:MR} shows the neutron star models constructed with FPS EOS, where the left panel corresponds to the stellar mass as a function of the central density normalized by $\varepsilon_s$, while the right panel corresponds to the stellar mass as a function of the stellar radius. In this figure, the solid line denotes the results in general relativity and the other lines denote the results in EiBI with various values of $8\pi\kappa\varepsilon_s$. From this figure, one can easily observe that the mass and radius of neutron stars depend strongly on the coupling constant in EiBI, even if EOS of neutron star matter is fixed. In fact, the stellar radii of $1.4M_\odot$ neutron stars in EiBI become $9.3$\% smaller for $8\pi\kappa\varepsilon_s=-0.02$, $7.6$\% larger for $8\pi\kappa\varepsilon_s=0.02$, and $16.5$\% larger for $8\pi\kappa\varepsilon_s=0.05$, compared with that in general relativity.

%%%%%%%%%%%%%%%%%%%%%%%%%%%%%%%%%%%
 %Figure 1
%%%%%%%%%%%%%%%%%%%%%%%%%%%%%%%%%%%
%%%%%%%%%%%%%%%%%%%%%%%%%%%%%%%%%%%

On such a neutron star model, we consider an axisymmetric magnetic field generated by a four currency $J^\mu$, adopting an ideal MHD approximation. The electromagnetic field is governed by the Maxwell equations with the physical metric $g_{\mu\nu}$,
\begin{gather}
  F_{[\mu\nu;\alpha]} = 0, \label{eq:Max1} \\
  F^{\mu\nu}_{\ \ ;\nu} = 4\pi J^\mu,  \label{eq:Max2}
\end{gather}
where $F_{\mu\nu}$ is the Faraday tensor and the covariant derivative would be calculated with the physical metric $g_{\mu\nu}$. Equation (\ref{eq:Max1}) automatically holds by introducing a vector potential, $A_\mu$, associated with $F_{\mu\nu}$ as $F_{\mu\nu}=A_{\nu,\mu}-A_{\mu,\nu}$. In order to determine the geometry of magnetic field, one also needs the equation of motions in addition to Eq. (\ref{eq:Max2}), which is obtained by projecting $T^{\mu\nu}_{\ \ ;\nu}=0$ on to the hypersurface normal to $u^\mu$. With the ideal MHD approximation, the equation of motions becomes
\begin{equation}
  (\varepsilon + p) u_{\mu;\nu}u^\nu + p_{,\mu} + u_\mu u^\nu p_{,\nu} = F_{\mu\nu}J^\nu.   \label{eq:motion}
\end{equation}
Now, assuming the appropriate gauge condition, $A_\mu$ can be described as $A_{\mu} = (0,A_r,0,A_\phi)$. In general, $A_\phi$ can be expanded, such as
\begin{equation}
  A_\phi(r,\theta) = a_\ell(r) \sin\theta\,\partial_\theta P_\ell(\cos\theta),  \label{eq:expand}
\end{equation}
where $P_\ell(\cos\theta)$ is the Legendre polynomial of the order $\ell$. Furthermore, we especially focus on the dipole magnetic field, i.e., $\ell=1$, because the dipole fields could be dominant in the neutron stars. Then, in the same way as in Refs. \cite{CFGP2008,SCK2008}, one can derive the equation to determine the vector potential $a_1$;
\begin{equation}
  a_1'' + \left(\frac{\nu'}{2}  - \frac{\lambda'}{2}\right)a_1' + \left(\zeta^2 e^{-\nu} - \frac{2}{f}\right)e^{\lambda}a_1
     = -4\pi e^{\lambda} j_1,  \label{eq:a1}
\end{equation}
where the prime denotes partial derivative with respect to $r$,  $j_1=c_0 f(\varepsilon+p)$, and $c_0$ is constant. We remark that constant $\zeta$ in Eq. (\ref{eq:a1}) is associated with the radial component of vector potential, i.e., $A_r = \zeta e^{-\nu/2+\lambda/2}a_\ell P_\ell$. The procedure how to derive Eq. (\ref{eq:a1}) is detailed in Appendix \ref{sec:appendix_1}.  Consequently, since the magnetic field can be given by $B_{\mu}=\varepsilon_{\mu\nu\alpha\beta}u^\nu F^{\alpha\beta}/2$, the components of the magnetic field $B_\mu$ are expressed as
\begin{gather}
  B_r = \frac{2a_1}{f} e^{\lambda/2} \cos\theta, \\
  B_\theta = - a_1' e^{-\lambda/2} \sin\theta, \\
  B_\phi = -\zeta a_1 e^{-\nu/2} \sin^2\theta,
\end{gather}
where $\varepsilon_{\mu\nu\alpha\beta}$ denotes the totally antisymmetric tensor and $\varepsilon_{tr\theta\phi}=\sqrt{-g}$. From these expressions, one can see that the constant $\zeta$ corresponds to the strength of toroidal magnetic field. Additionally, the tetrad components of magnetic files are given by 
\begin{gather}
  B_{[r]} = 2a_1 f^{-1}\cos\theta, \label{eq:Br} \\
  B_{[\theta]} = -a_1' f^{-1/2} e^{-\lambda/2}\sin\theta,  \label{eq:Bt} \\
  B_{[\phi]} = -\zeta a_1 f^{-1/2} e^{-\nu/2}\sin\theta.  \label{eq:Bp}
\end{gather}

Since we consider that the exterior region of the star is in vacuum, as mentioned before, the spacetime outside the star becomes the same as that in general relativity, which can be described as the Schwarzschild metric. In such spacetime, the poloidal magnetic field ($\zeta=0$) is analytically given by 
\begin{equation}
  a_1^{\rm (ex)} = -\frac{3\mu_br^2}{8M^3}\left[\ln\left(1-\frac{2M}{r}\right) + \frac{2M}{r} + \frac{2M^2}{r^2}\right], \label{eq:ex}
\end{equation}
where $\mu_b$ is the magnetic dipole moment observed at infinity \cite{WS1983}. Thus, at the stellar surface, the interior solution determined from Eq. (\ref{eq:a1}) should be connected to the exterior solution [Eq. (\ref{eq:ex})] in such a way that $a_1$ and $a_1'$ become continuous. In practice, from Eq. (\ref{eq:a1}), one can show that the behavior of $a_1$ in the vicinity of the stellar center is expressed as $a_1=\alpha_0r^2 + {\cal O}(r^4)$, where $\alpha_0$ is an arbitrary constant. So, the arbitrary constants $\alpha_0$ and $c_0$, which is a constant in the four currency $j_1$, are determined so that $a_1$ and $a_1'$ should be continuous at the stellar surface.

%%%%%%%%%%%%%%%%%%%%%%%%%%%%%%%%%%%%%%%%%%%%%%%%
\section{Numerical Results}
\label{sec:III}
%%%%%%%%%%%%%%%%%%%%%%%%%%%%%%%%%%%%%%%%%%%%%%%%

The magnetic field strength $B$ is calculated by $B=(B_\mu B_\nu g^{\mu\nu})^{1/2}$, which can be expressed as
\begin{equation}
  B = f^{-1}\left[4a_1^2\cos^2\theta + a_1'^2 f e^{-\lambda} \sin^2\theta + \zeta^2 a_1^2 f e^{-\nu} \sin^2\theta\right]^{1/2}.
\end{equation}
Thus, one can show that the magnetic field strength at the stellar center is $B_0 = 2\alpha_0a_0b_0$, where $a_0=\sqrt{1+8\pi\kappa\varepsilon_0}$ and $b_0=\sqrt{1-8\pi\kappa p_0}$, while $\varepsilon_0$ and $p_0$ denote the central values of $\varepsilon$ and $p$. 
%We remark that $\zeta^{-1}$ is a parameter with the dimension of length.
%We remark that the central strength of magnetic field is independent from the existence of the toroidal magnetic component $B_\phi$, i.e., $B_0$ is independent from $\zeta$.
In the limit of $\kappa=0$, this expression reduces to that in general relativity \cite{ST2015}. The concrete structure of magnetic fields is discussed below, where we separately examine the pure poloidal magnetic fields ($\zeta=0$) in \S\ref{sec:IIIa} and the mixed magnetic fields ($\zeta\ne 0$) in \S\ref{sec:IIIb}.

%%%%%%%%%%%%%%%%%%%%%%%%%%%%%%%%%%%%%%%%%%%%%%%%
\subsection{Pure Poloidal Magnetic Fields ($\zeta=0$)}
\label{sec:IIIa}
%%%%%%%%%%%%%%%%%%%%%%%%%%%%%%%%%%%%%%%%%%%%%%%%

\begin{figure*}
\begin{center}
\begin{tabular}{cc}
\includegraphics[scale=0.5]{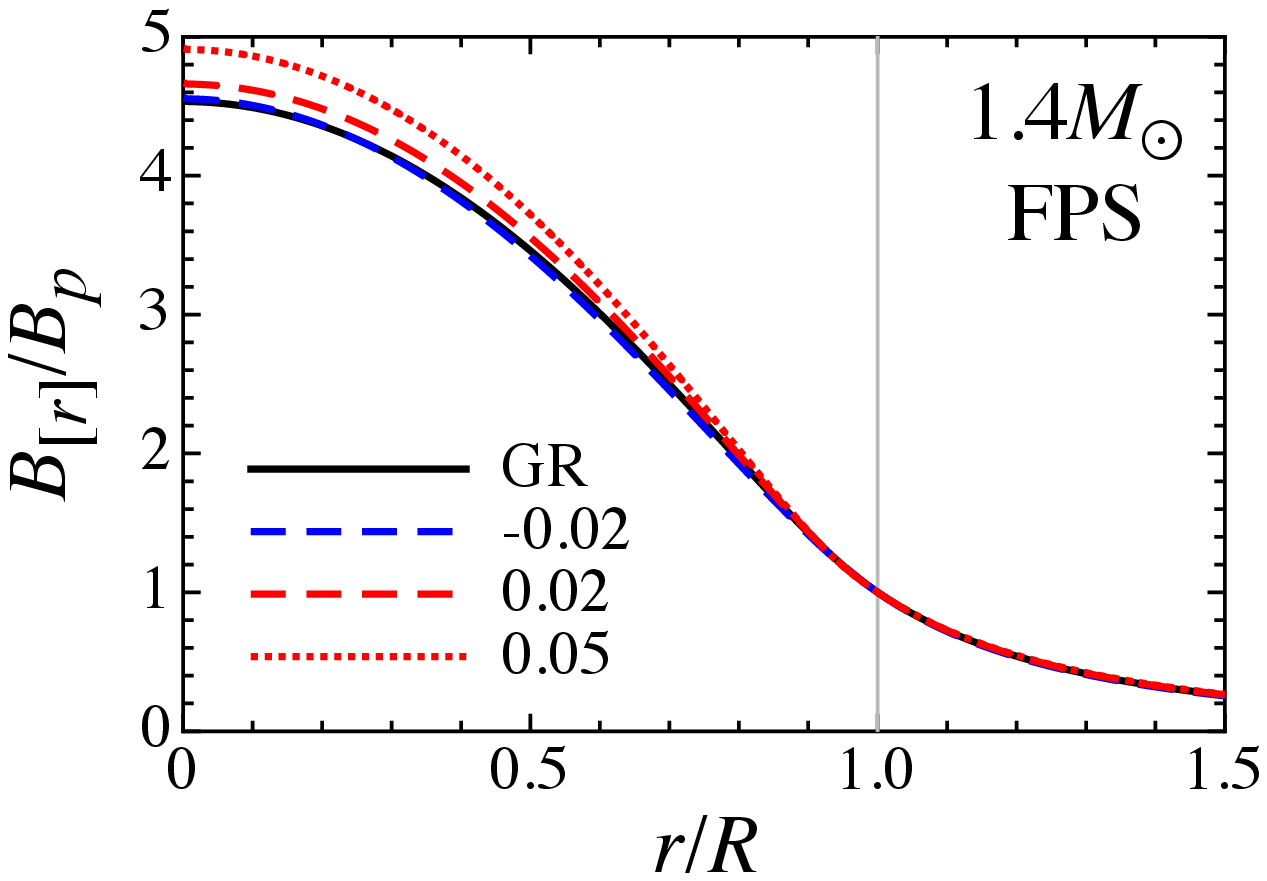} &
\includegraphics[scale=0.5]{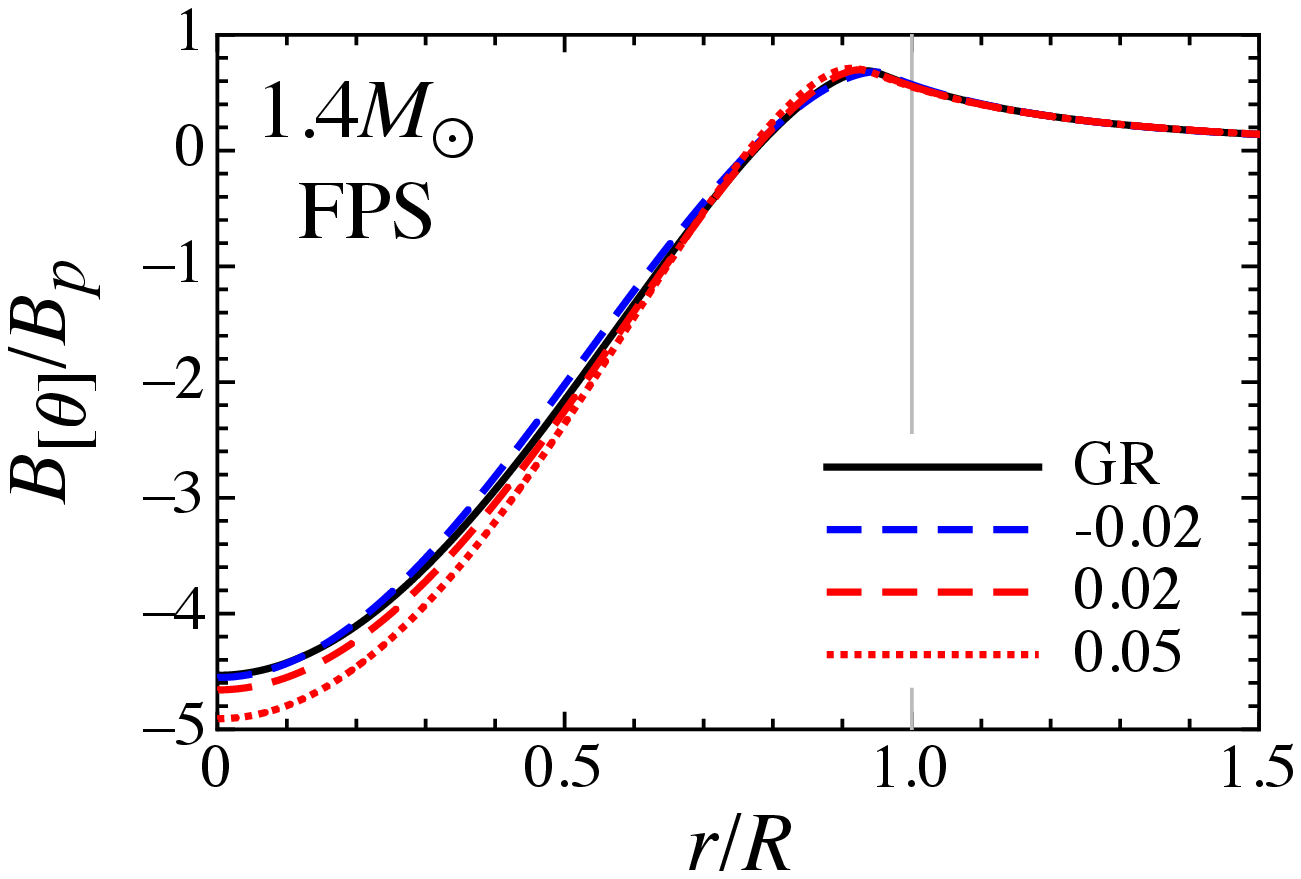}
\end{tabular}
\end{center}
\caption{%%
For the stellar models with $M=1.4M_\odot$ for FPS EOS, the tetrad components of the pure poloidal magnetic fields are plotted as a function of $r/R$, where the left and right panels correspond to the radial component on the symmetry axis ($\theta=0$) and the $\theta$-component on the equatorial plane ($\theta=\pi/2$), respectively. The both components are normalized by $B_p$, which is the magnetic field strength at the stellar surface of the poles. The solid line corresponds to the result in general relativity, while the broken and dotted lines correspond to the results in EiBI with various values of $8\pi\kappa\varepsilon_s$. The vertical lines denote the position of the stellar surface. 
}
\label{fig:BM14}
\end{figure*}

\begin{figure*}
\begin{center}
\includegraphics[scale=0.5]{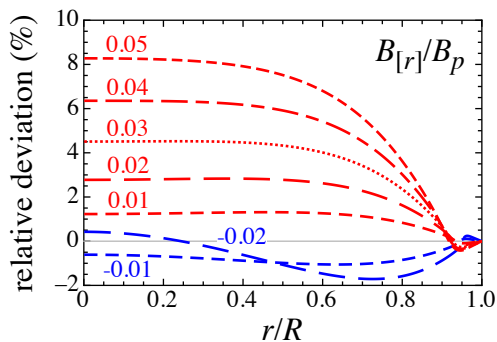}
\end{center}
\caption{%%
Relative deviation of $B_{[r]}/B_p$ in EiBI from that in general relativity for the stellar models with $M=1.4M_\odot$ constructed with FPS EOS, which is corresponding to the left panel in Fig. \ref{fig:BM14}. The labels in the figure denote the values of coupling constant in EiBI.
}
\label{fig:deviation}
\end{figure*}

\begin{figure*}
\begin{center}
\begin{tabular}{ccc}
\includegraphics[scale=0.5]{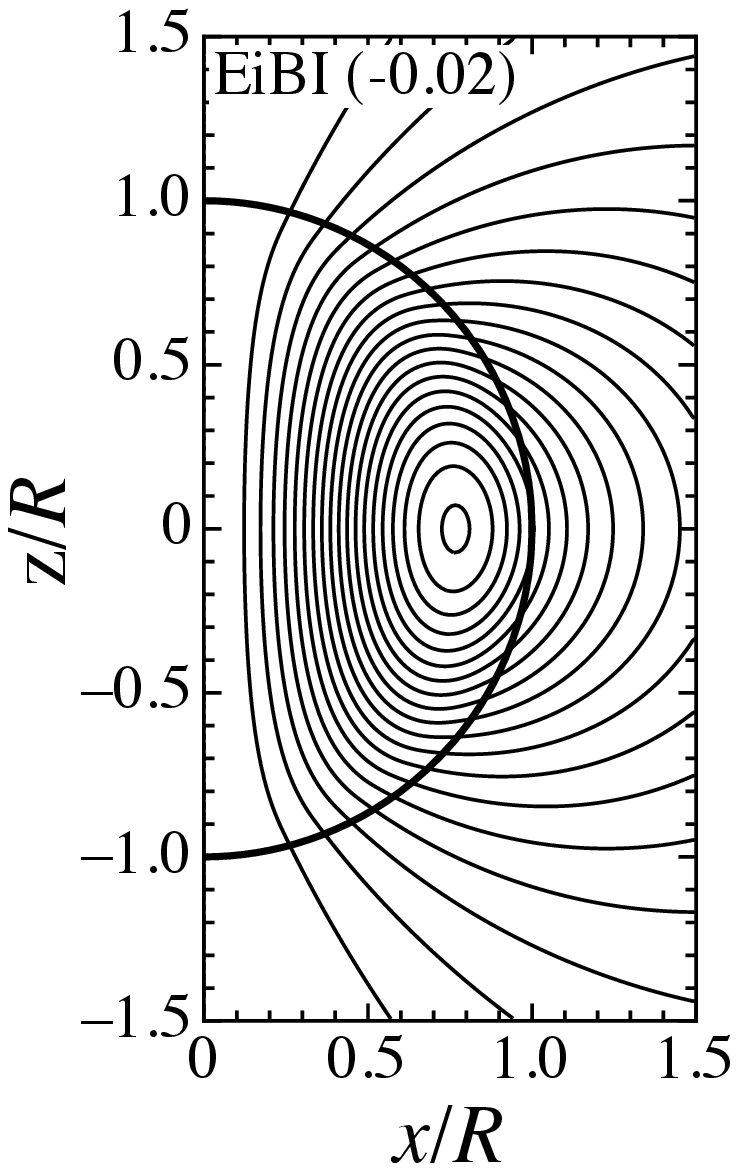} &
\includegraphics[scale=0.5]{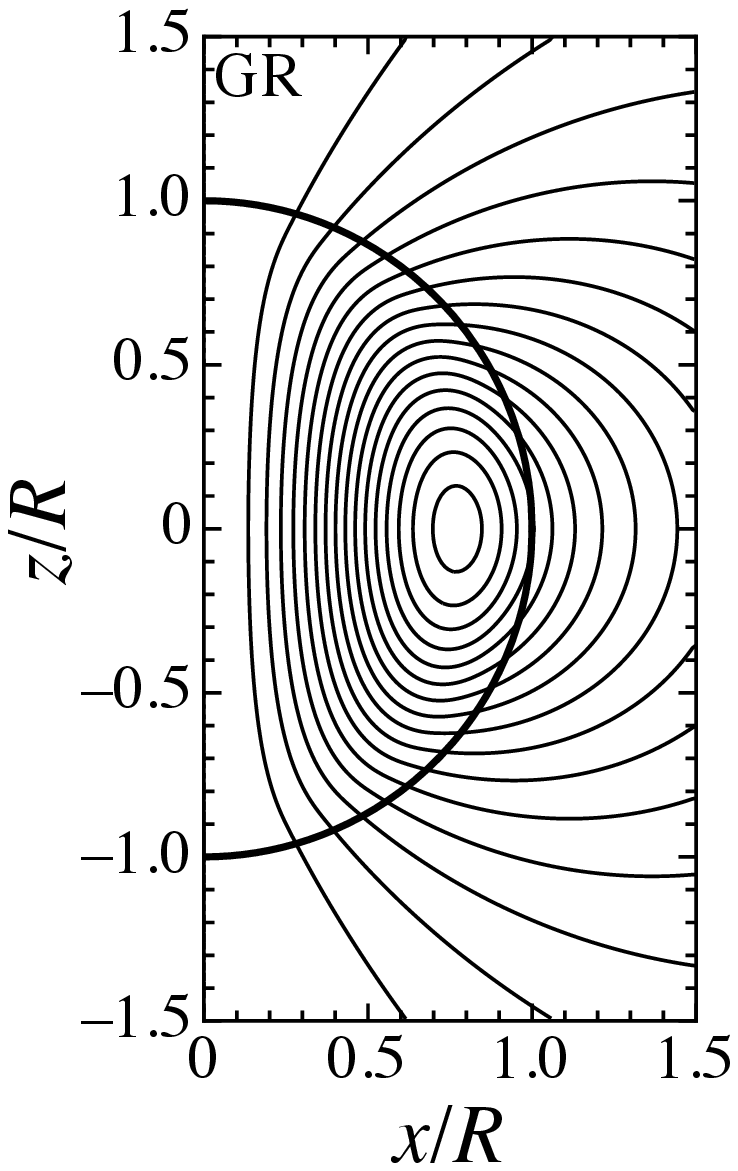} &
\includegraphics[scale=0.5]{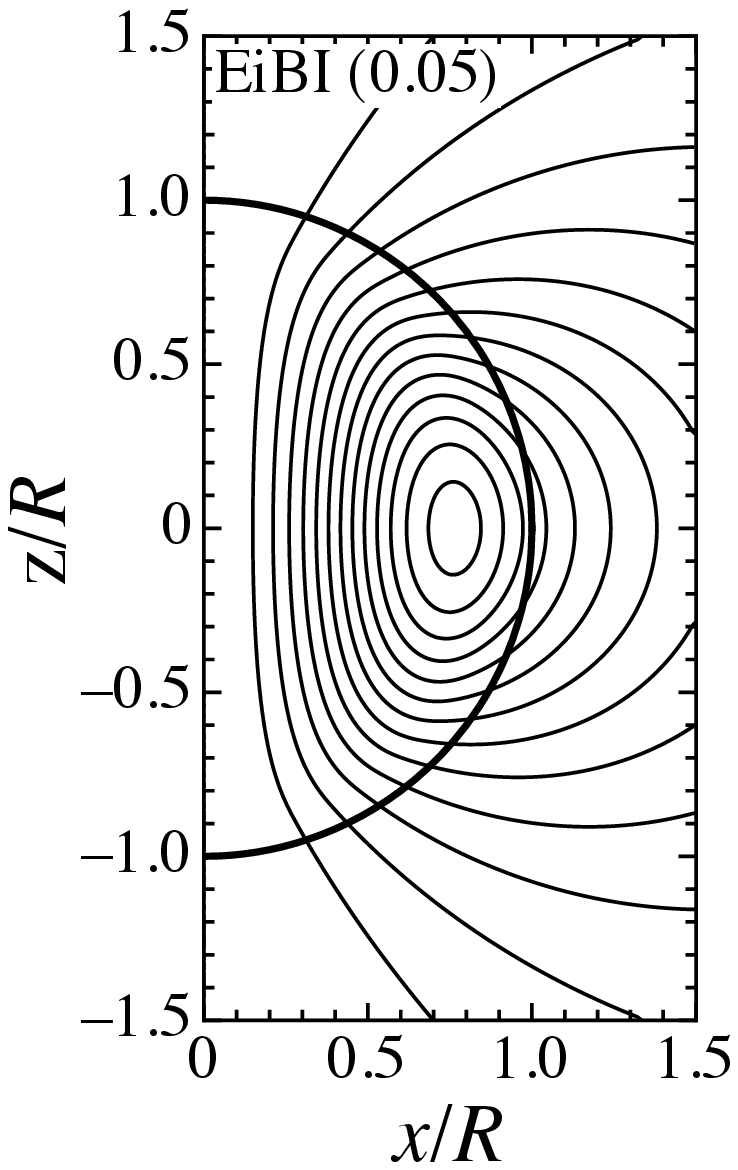}
\end{tabular}
\end{center}
\caption{%%
Magnetic configurations on the meridional plane for the stellar models with $M=1.4M_\odot$ for FPS EOS in GR (middle panel) and in EiBI with $8\pi\kappa\varepsilon_s=-0.02$ (left panel) and 0.05 (right panel). The magnetic field strength is normalized by the magnetic dipole moment $\mu_b$.
}
\label{fig:Bcon}
\end{figure*}

\begin{figure*}
\begin{center}
\begin{tabular}{cc}
\includegraphics[scale=0.5]{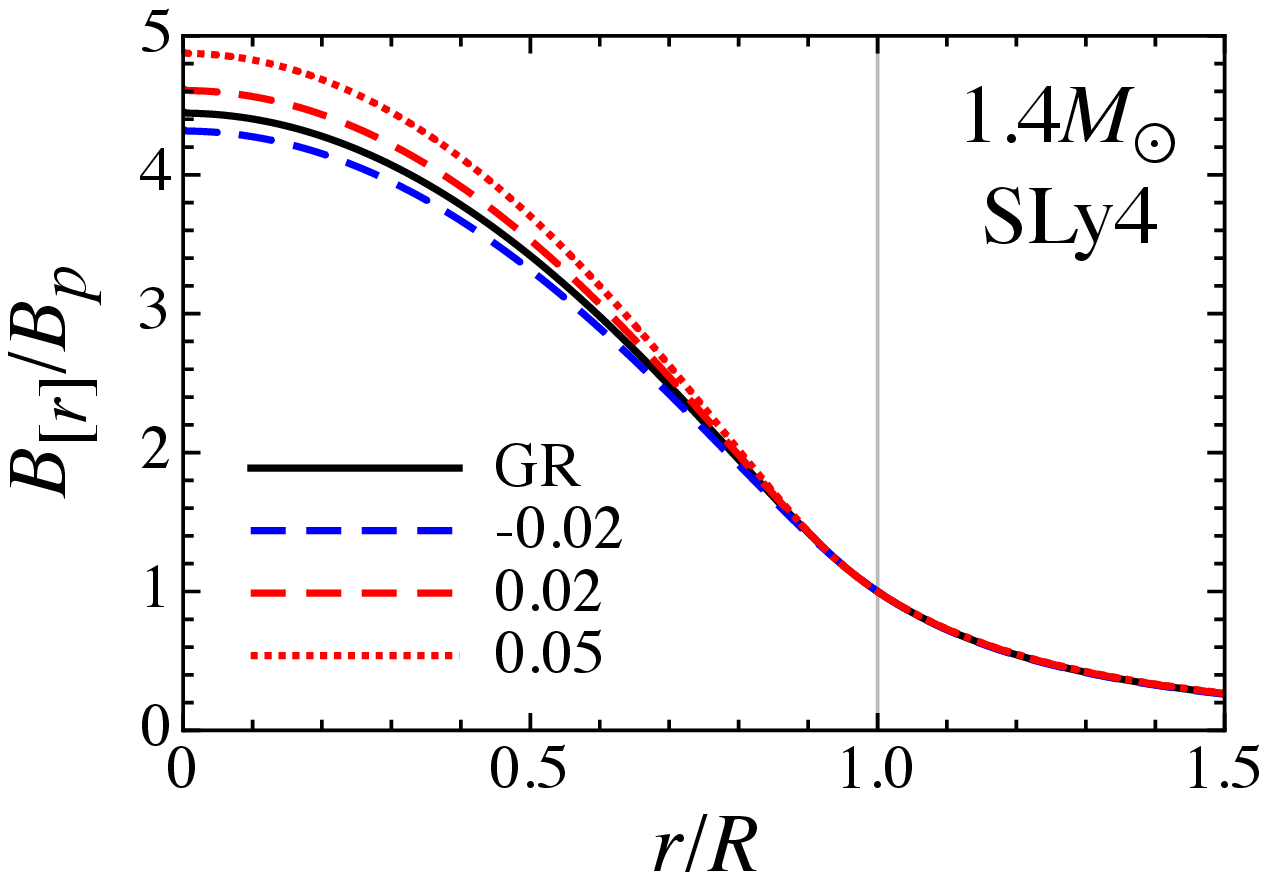} &
\includegraphics[scale=0.5]{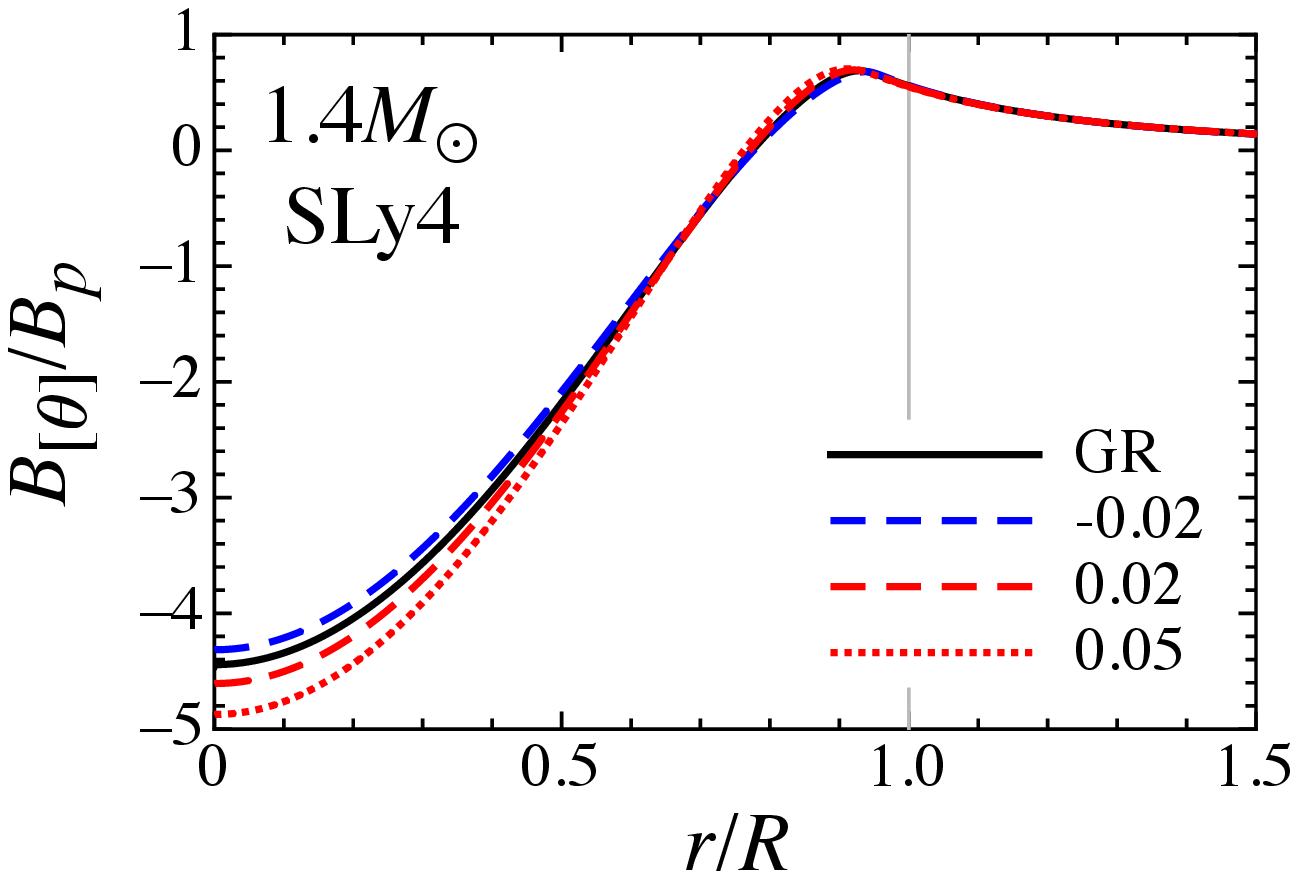}
\end{tabular}
\end{center}
\caption{%%
Similar to Fig. \ref{fig:BM14}, but for the stellar models constructed with SLy4 EOS.
}
\label{fig:BM14s}
\end{figure*}

First, one can show that the magnetic distribution is scaled by the magnetic field strength at the stellar surface of the poles ($\theta=0$), $B_p$, if the stellar model is fixed. That is, the distributions of $B_{[i]}/B_p$ for $i=r$, $\theta$, and $\phi$ are independent of $B_p$ for each stellar model. In Fig. \ref{fig:BM14}, we show the distributions of $B_{[r]}/B_p$ on the symmetry axis ($\theta=0$) in the left panel and $B_{[\theta]}/B_p$ on the equatorial plane ($\theta=\pi/2$) in the right panel for the stellar models with $M=1.4M_\odot$ contracted with FPS EOS, where the solid line corresponds to the result in general relativity and the other lines correspond to the results in EiBI with various values of $8\pi\kappa\varepsilon_s$. From this figure, one can observe that the magnetic distributions in EiBI are qualitatively the same as that in general relativity. In fact, the deviation between the results in general relativity and in EiBI is not so much. In Fig. \ref{fig:deviation}, we show the relative deviation of $B_{[r]}/B_p$ in EiBI from that in general relativity for the stellar models with $M=1.4M_\odot$ constructed with FPS EOS. From this figure, we find that the deviation from general relativity is at most $10$ \% with the coupling constant in EiBI adopted in this paper. In particular, the magnetic distribution in the crust region depends weakly on the coupling constant in EiBI, which is less than $0.5\%$. That is, apart from the gravitational theory, one might be able to discuss the magnetic properties in the crust region of neutron stars. In addition, we show the magnetic configurations on the meridional plane for the stellar models with $M=1.4M_\odot$ for FPS EOS in Fig. \ref{fig:Bcon}, where the middle panel corresponds to that in GR ($\kappa=0)$, while the left and right panels correspond to those in EiBI with $8\pi\kappa\varepsilon_s=-0.02$ and 0.05. The magnetic field strength is normalized by the magnetic dimple moment. As shown in Fig. \ref{fig:BM14}, the magnetic configurations in EiBI are quite similar to that in GR. As with Fig. \ref{fig:BM14}, we also show the magnetic distributions for the stellar models with $M=1.4M_\odot$ constructed with SLy4 EOS in Fig. \ref{fig:BM14s}. Comparing Fig. \ref{fig:BM14} with Fig. \ref{fig:BM14s}, we find that the dependence of magnetic distribution on the coupling constant in EiBI is comparable to that on EOS for neutron star matter.

%%%%%%%%%%%%%%%%%%%%%%%%%%%%%%%%%%%
 %Figure 2
%%%%%%%%%%%%%%%%%%%%%%%%%%%%%%%%%%%
%%%%%%%%%%%%%%%%%%%%%%%%%%%%%%%%%%%

%%%%%%%%%%%%%%%%%%%%%%%%%%%%%%%%%%%
 %Figure 2a
%%%%%%%%%%%%%%%%%%%%%%%%%%%%%%%%%%%
%%%%%%%%%%%%%%%%%%%%%%%%%%%%%%%%%%%

%%%%%%%%%%%%%%%%%%%%%%%%%%%%%%%%%%%
 %Figure 2b
%%%%%%%%%%%%%%%%%%%%%%%%%%%%%%%%%%%
%%%%%%%%%%%%%%%%%%%%%%%%%%%%%%%%%%%

%%%%%%%%%%%%%%%%%%%%%%%%%%%%%%%%%%%
 %Figure 3
%%%%%%%%%%%%%%%%%%%%%%%%%%%%%%%%%%%
%\begin{figure*}
%\begin{center}
%\begin{tabular}{cc}
%\includegraphics[scale=0.5]{Br-M16a} &
%\includegraphics[scale=0.5]{Bt-M16a}
%\end{tabular}
%\end{center}
%\caption{%%
%Similar to Fig. \ref{fig:BM14}, but for the stellar model with $M=1.6M_\odot$.
%}
%\label{fig:BM16}
%\end{figure*}
%%%%%%%%%%%%%%%%%%%%%%%%%%%%%%%%%%%

%%%%%%%%%%%%%%%%%%%%%%%%%%%%%%%%%%%
 %Figure 3
%%%%%%%%%%%%%%%%%%%%%%%%%%%%%%%%%%%
%%%%%%%%%%%%%%%%%%%%%%%%%%%%%%%%%%%

\begin{figure*}
\begin{center}
\begin{tabular}{cc}
\includegraphics[scale=0.5]{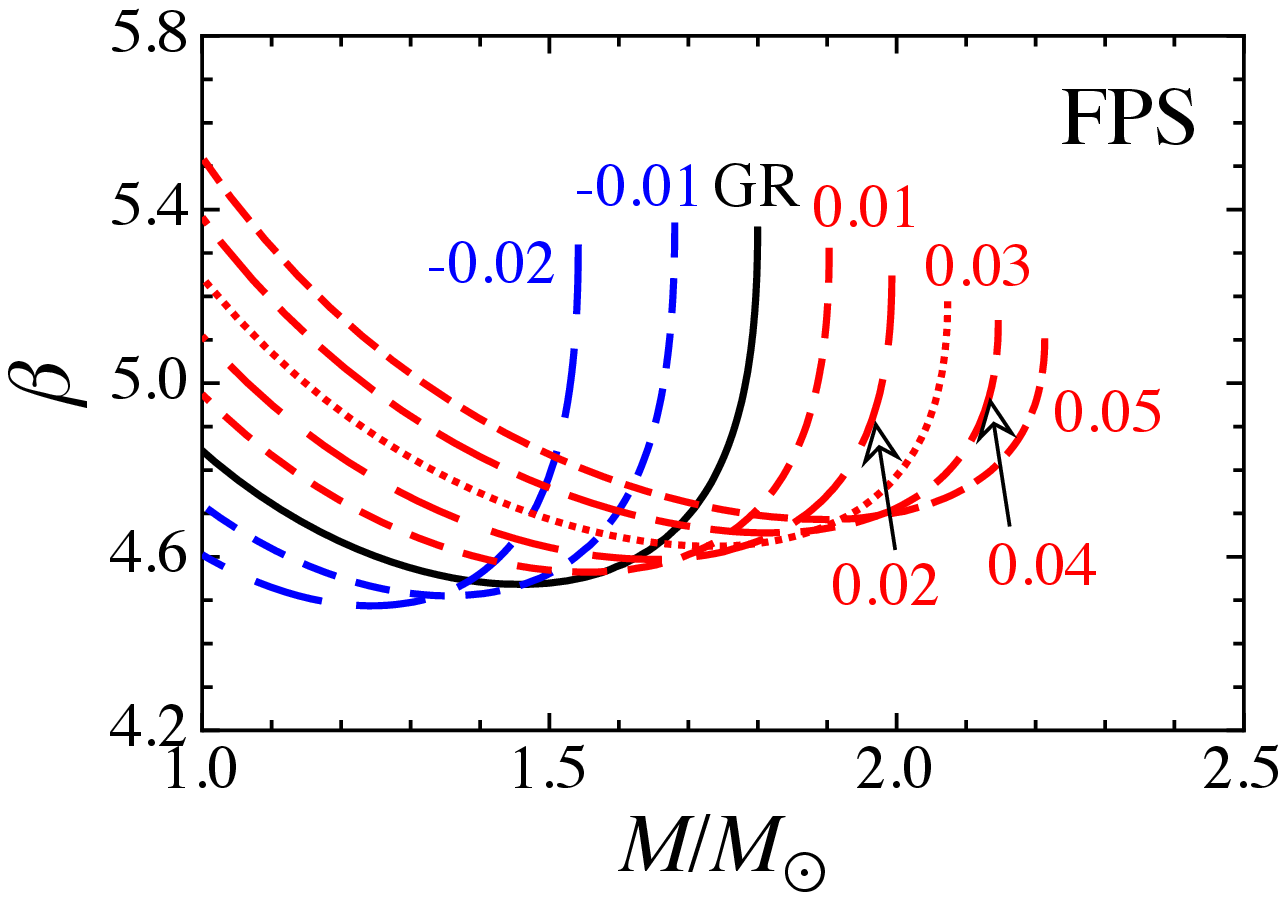} &
\includegraphics[scale=0.5]{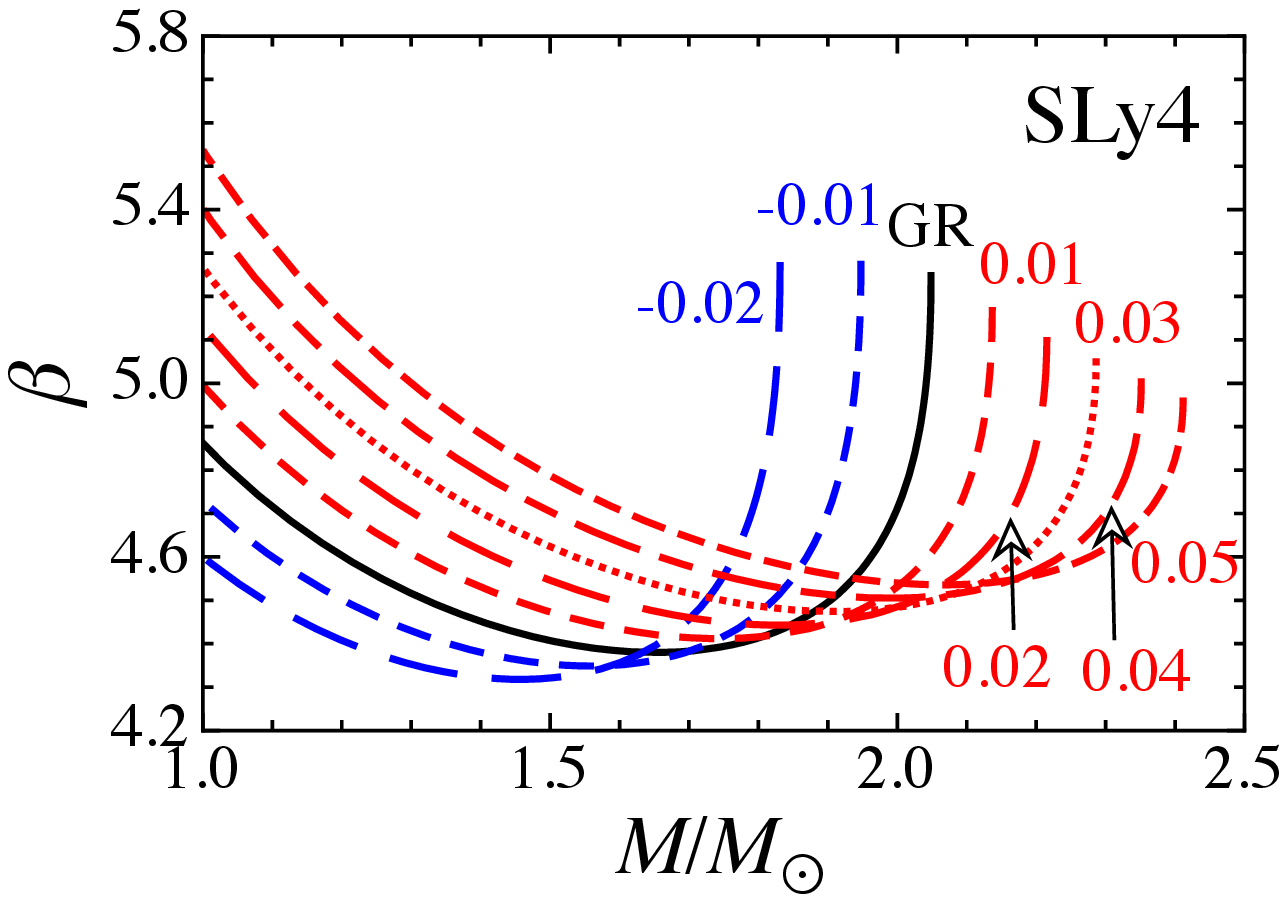}
\end{tabular}
\end{center}
\caption{%%
The proportionality factor $\beta$ for the stellar models with various values of coupling constant in EiBI are shown as a function of the stellar mass, where the left and right panels correspond to the results for FPS and SLy4 EOSs, respectively. The labels in the figure denote the values of coupling constant in EiBI.
}
\label{fig:beta}
\end{figure*}

\begin{figure*}
\begin{center}
\begin{tabular}{cc}
\includegraphics[scale=0.5]{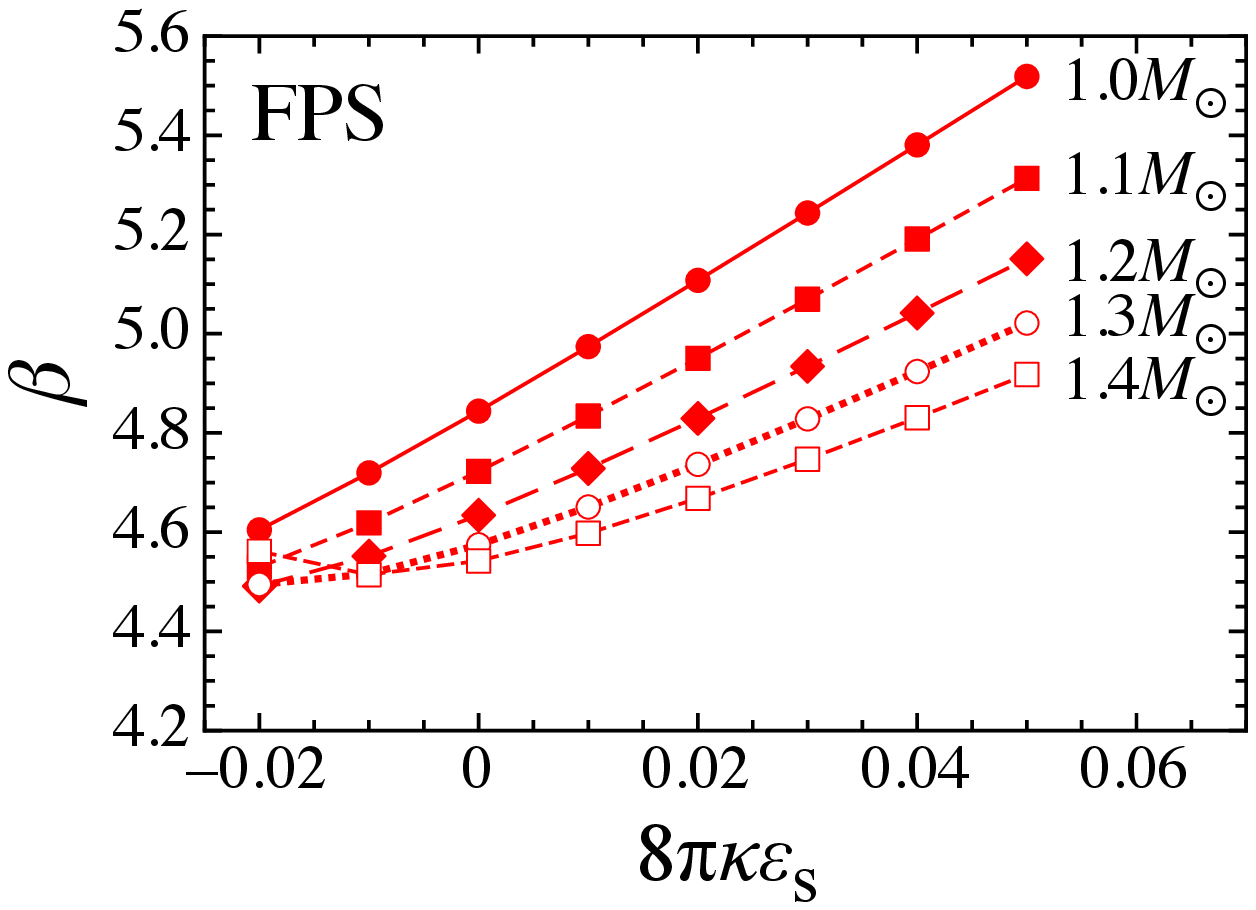} &
\includegraphics[scale=0.5]{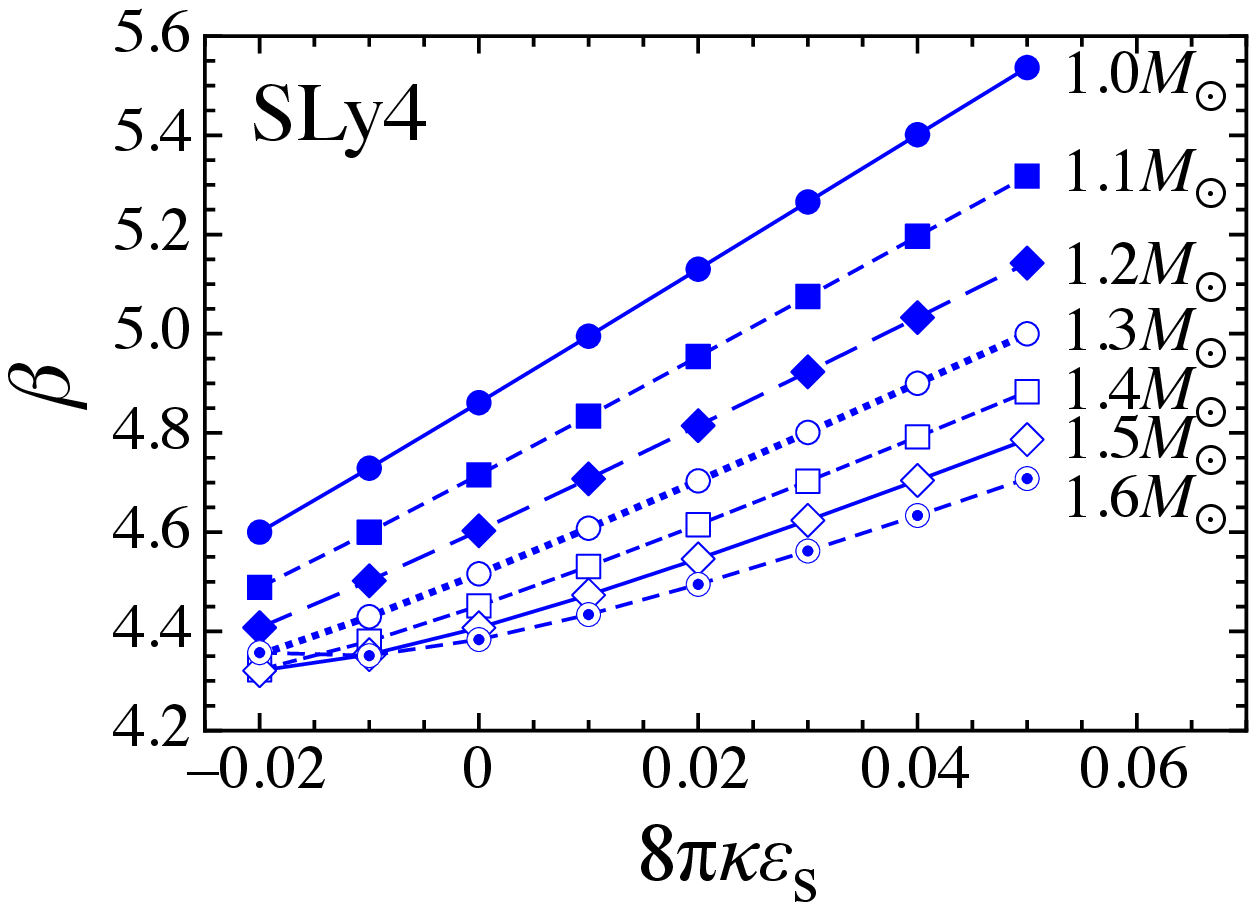}
\end{tabular}
\end{center}
\caption{%%
The proportionality factor $\beta$ as a function of the coupling constant in EiBI for the fixed stellar mass, where the left and right panels correspond to the results for FPS and SLy4 EOSs, respectively.
}
\label{fig:Mbeta}
\end{figure*}

Moreover, the magnetic field strength at the stellar center can be also scaled by $B_p$ for each stellar model, such as
\begin{equation}
  B_0 = \beta B_p, \label{eq:beta}
\end{equation}
where $\beta$ is a proportionality constant \cite{ST2015}. In Fig. \ref{fig:beta}, we show the proportionality factor $\beta$ as a function of the stellar mass with various values of the coupling constant in EiBI, where the left and right panels correspond to the results for the stellar models constructed with FPS and SLy4 EOSs, respectively. From this figure, one can see that the value of $\beta$ is almost independent of the adopted EOS for neutron star matter, which is $\sim 5$. Additionally, for the stellar models whose masses are smaller than a critical value depending on the adopted EOS, $\beta$ for the fixed stellar mass is almost proportional to the coupling constant in EiBI at least in the range adopted in this paper. This statement is clear from Fig. \ref{fig:Mbeta}, where the proportionality factor in Eq. (\ref{eq:beta}) for the fixed stellar mass is shown as a function of the coupling constant in EiBI. From this figure, such a critical stellar mass would be around $1.2M_\odot$ for FPS EOS and $1.4M_\odot$ for SLy4 EOS. Furthermore, from this figure, we find that $\beta$ for each coupling constant $8\pi\kappa\varepsilon_s$ depends weakly on the EOS for neutron star matter, if the mass of neutron star would be very low, for instance $M\simeq M_\odot$. 
%This must be another feature for low-mass neutron star, as well as the relation between the radius of $0.5M_\odot$ neutron star and the neutron skin thickness of ${}^{208}$Pb is almost independent of the adopted EOS \cite{S2014a}.

%%%%%%%%%%%%%%%%%%%%%%%%%%%%%%%%%%%
 %Figure 5
%%%%%%%%%%%%%%%%%%%%%%%%%%%%%%%%%%%
%%%%%%%%%%%%%%%%%%%%%%%%%%%%%%%%%%%

%%%%%%%%%%%%%%%%%%%%%%%%%%%%%%%%%%%
 %Figure 6
%%%%%%%%%%%%%%%%%%%%%%%%%%%%%%%%%%%
%%%%%%%%%%%%%%%%%%%%%%%%%%%%%%%%%%%

%%%%%%%%%%%%%%%%%%%%%%%%%%%%%%%%%%%%%%%%%%%%%%%%
\subsection{Mixed Magnetic Fields ($\zeta\ne0$)}
\label{sec:IIIb}
%%%%%%%%%%%%%%%%%%%%%%%%%%%%%%%%%%%%%%%%%%%%%%%%

\begin{figure*}
\begin{center}
\begin{tabular}{ccc}
\includegraphics[scale=0.42]{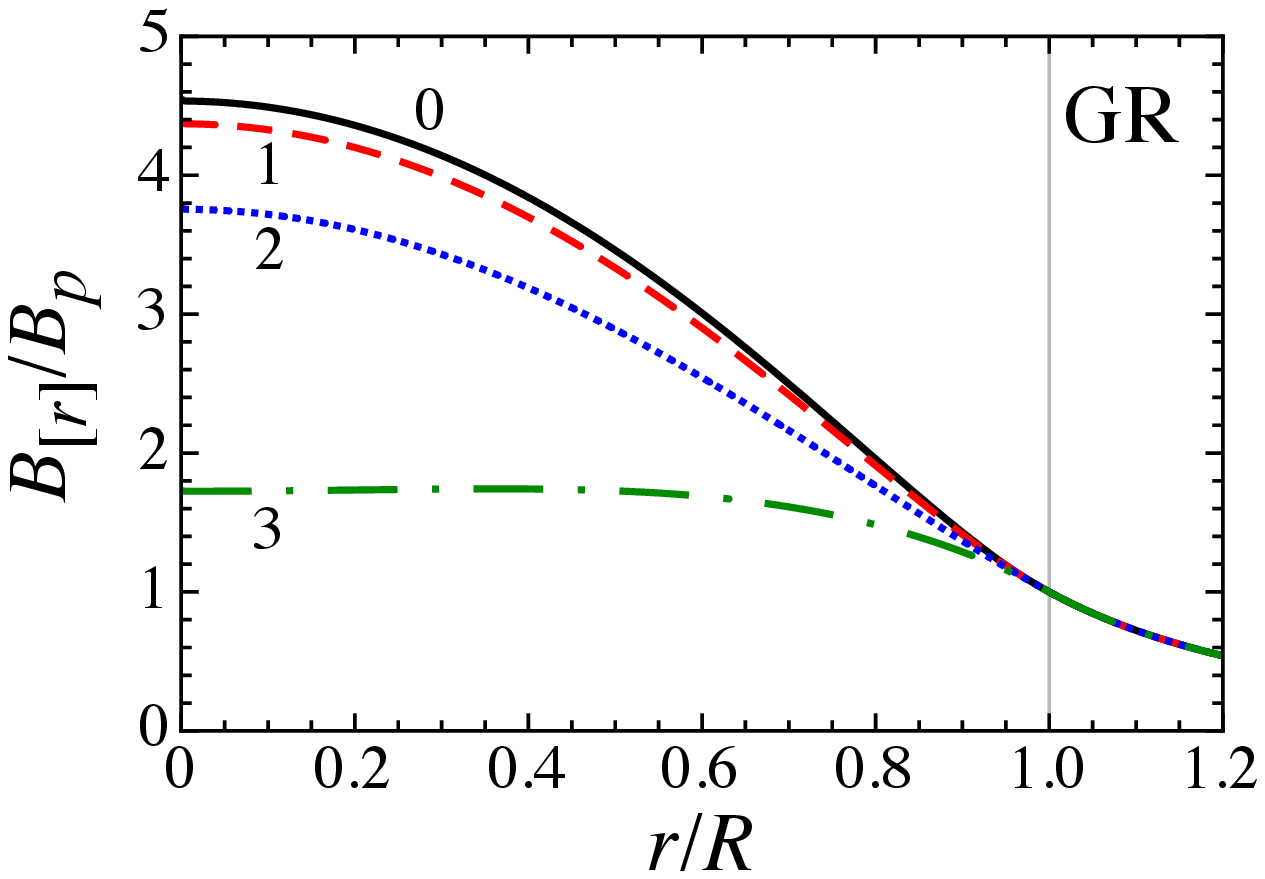} & \includegraphics[scale=0.42]{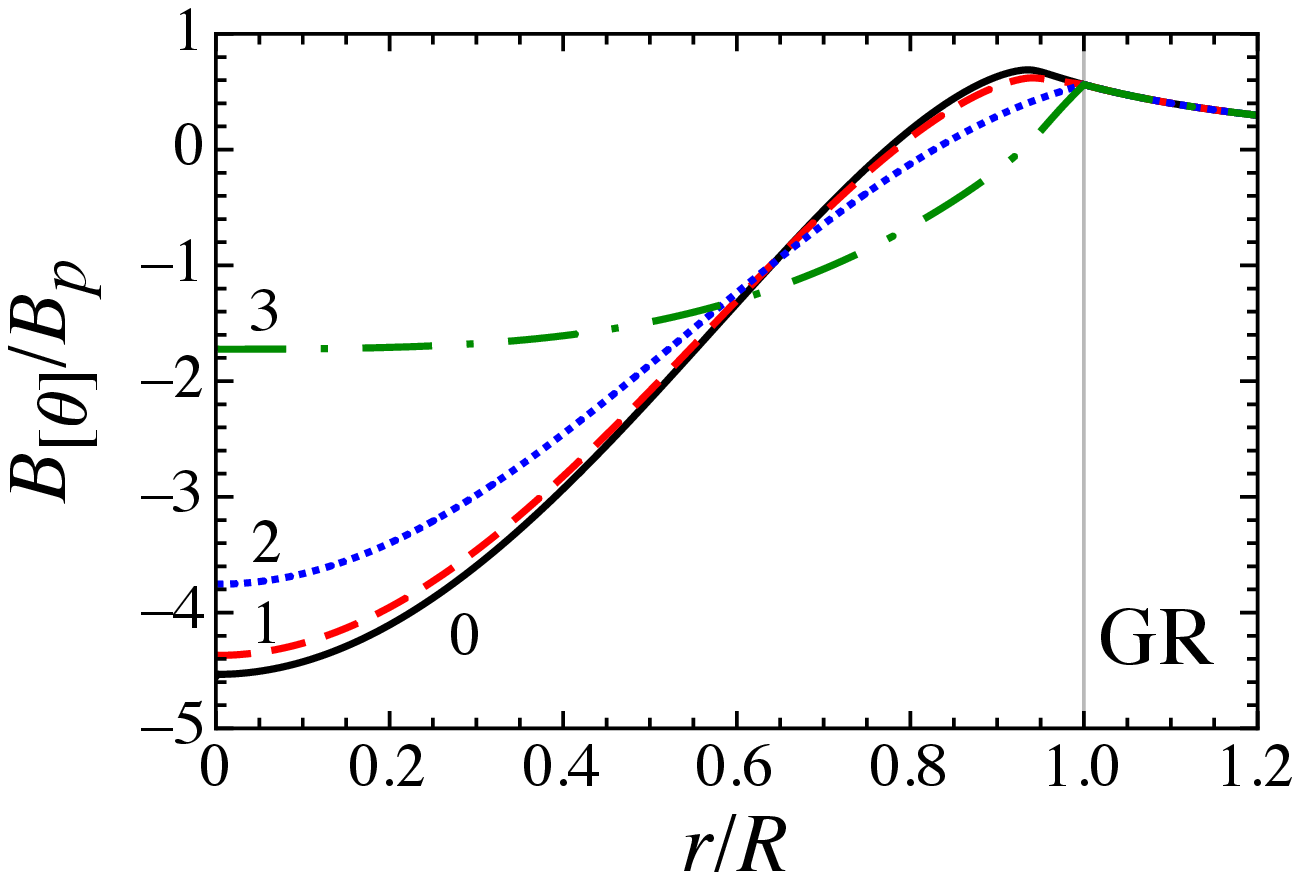} & \includegraphics[scale=0.42]{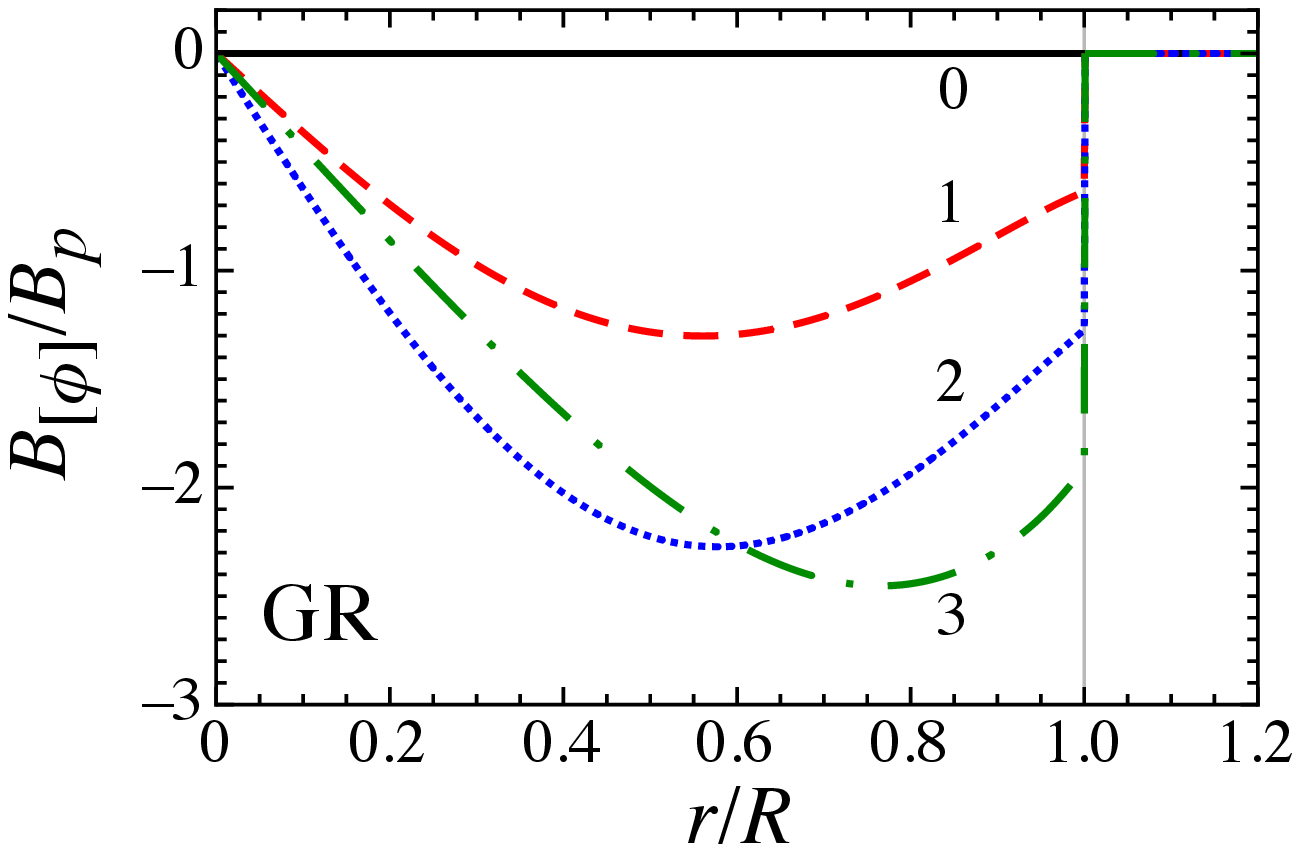}
\end{tabular}
\end{center}
\caption{%%
For the stellar models in general relativity with $M=1.4M_\odot$ constructed with FPS EOS, the tetrad components of the mixed magnetic fields normalized by $B_p$ are plotted as a function of $r/R$, where the left, middle, and right panels are $B_{[r]}/B_p$ on the symmetry axis ($\theta=0$), $B_{[\theta]}/B_p$ on the equatorial plane ($\theta=\pi/2$), and $B_{[\phi]}/B_p$ on the equatorial plane, respectively. The different lines in the figure correspond to the magnetic field distributions with different values of $\zeta R$, and the labels in the figure denote the value of $\zeta R$.
}
\label{fig:BM14T0}
\end{figure*}

As with the case of the pure poloidal magnetic fields shown in the previous subsection, the distribution of the mixed magnetic fields is also scaled by $B_p$, and the profiles of $B_{[i]}/B_p$ for $i=r$, $\theta$, and $\phi$ are independent of the strength of $B_p$ for each stellar model. For reference, first, we show the magnetic distributions in general relativity for the stellar models with $M=1.4M_\odot$ constructed with FPS EOS in Fig. \ref{fig:BM14T0}. In this figure, the left, middle, and right panels correspond to $B_{[r]}/B_p$ on the symmetry axis ($\theta=0$), $B_{[\theta]}/B_p$ on the equatorial plane ($\theta=\pi/2$), and $B_{[\phi]}/B_p$ on the equatorial plane, respectively. The solid line denotes the magnetic distribution for the pure poloidal field, while the other lines denote those for the mixed fields. As mentioned before, the toroidal magnetic component is characterized by the parameter $\zeta$, such as Eq. (\ref{eq:Bp}). In Fig. \ref{fig:BM14T0}, we show the magnetic distributions with the variable values of $\zeta$ normalized by $1/R$, because $\zeta$ is a parameter with the dimension of inverse of length and then $\zeta R$ becomes a dimensionless parameter. From this figure, one can observe that the distributions of $B_{[r]}/B_p$ and $B_{[\theta]}/B_p$ are also changed due to the existence of the toroidal magnetic field, where the central field strengths of $B_{[r]}/B_p$ and $B_{[\theta]}/B_p$ decrease with the value of $\zeta$. This result suggests the existence of the maximum of $\zeta$, where the central values of $B_{[r]}/B_p$ and $B_{[\theta]}/B_p$ become zero. Hereafter, such a maximum of $\zeta$ denotes $\zeta_{\rm max}$, and  $\zeta_{\rm max}R=3.30$ for the case of the neutron star model in Fig. \ref{fig:BM14T0}. In practice, with $\zeta$ more than $\zeta_{\rm max}$, the direction of the magnetic field can be opposite inside the star \cite{CFGP2008,SCK2008}. Additionally, from Fig. \ref{fig:BM14T0}, one can see that the position where $|B_{[\phi]}/B_p|$ becomes maximum is shifting outward with the value of $\zeta$.

%%%%%%%%%%%%%%%%%%%%%%%%%%%%%%%%%%%
 %Figure 7
%%%%%%%%%%%%%%%%%%%%%%%%%%%%%%%%%%%
%%%%%%%%%%%%%%%%%%%%%%%%%%%%%%%%%%%

\begin{figure*}
\begin{center}
\begin{tabular}{ccc}
\includegraphics[scale=0.42]{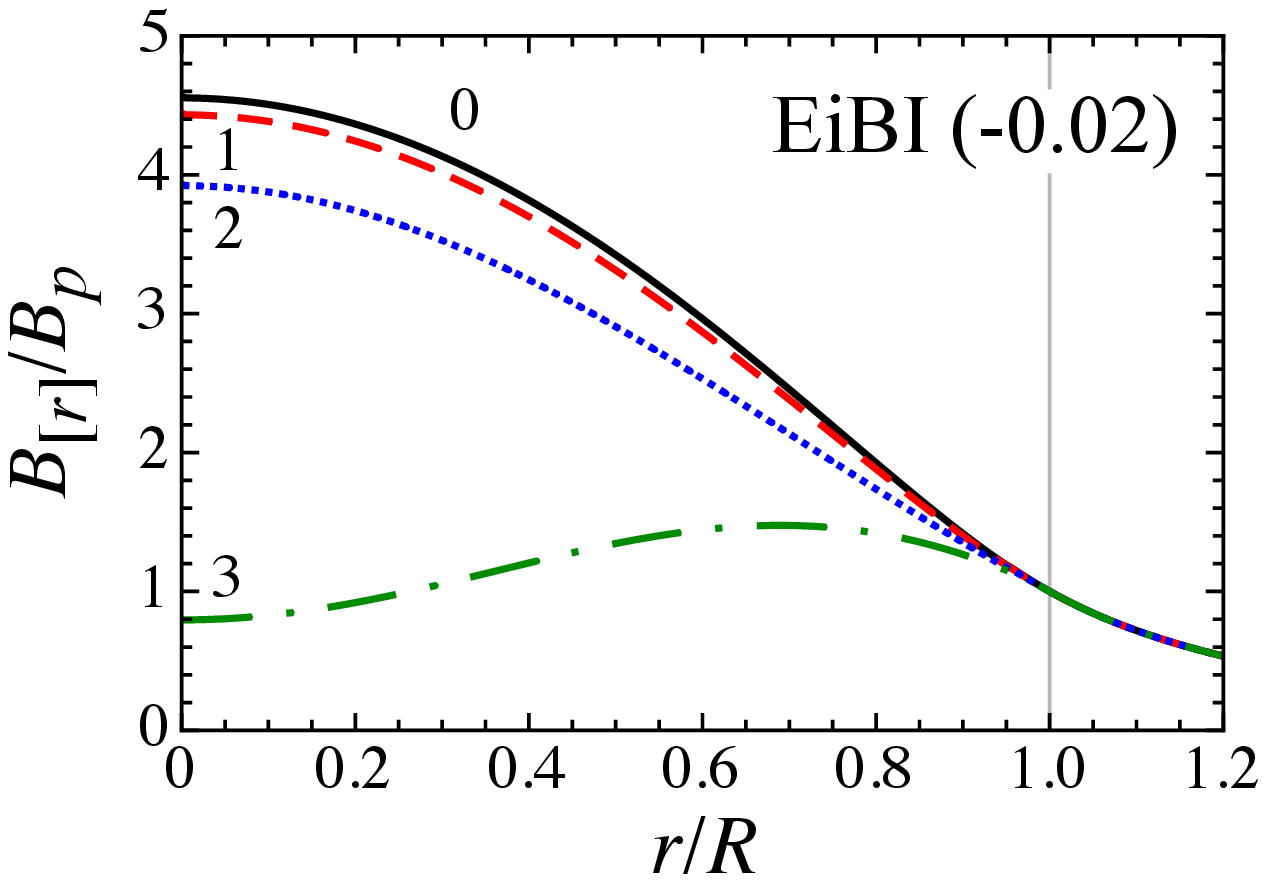} & \includegraphics[scale=0.42]{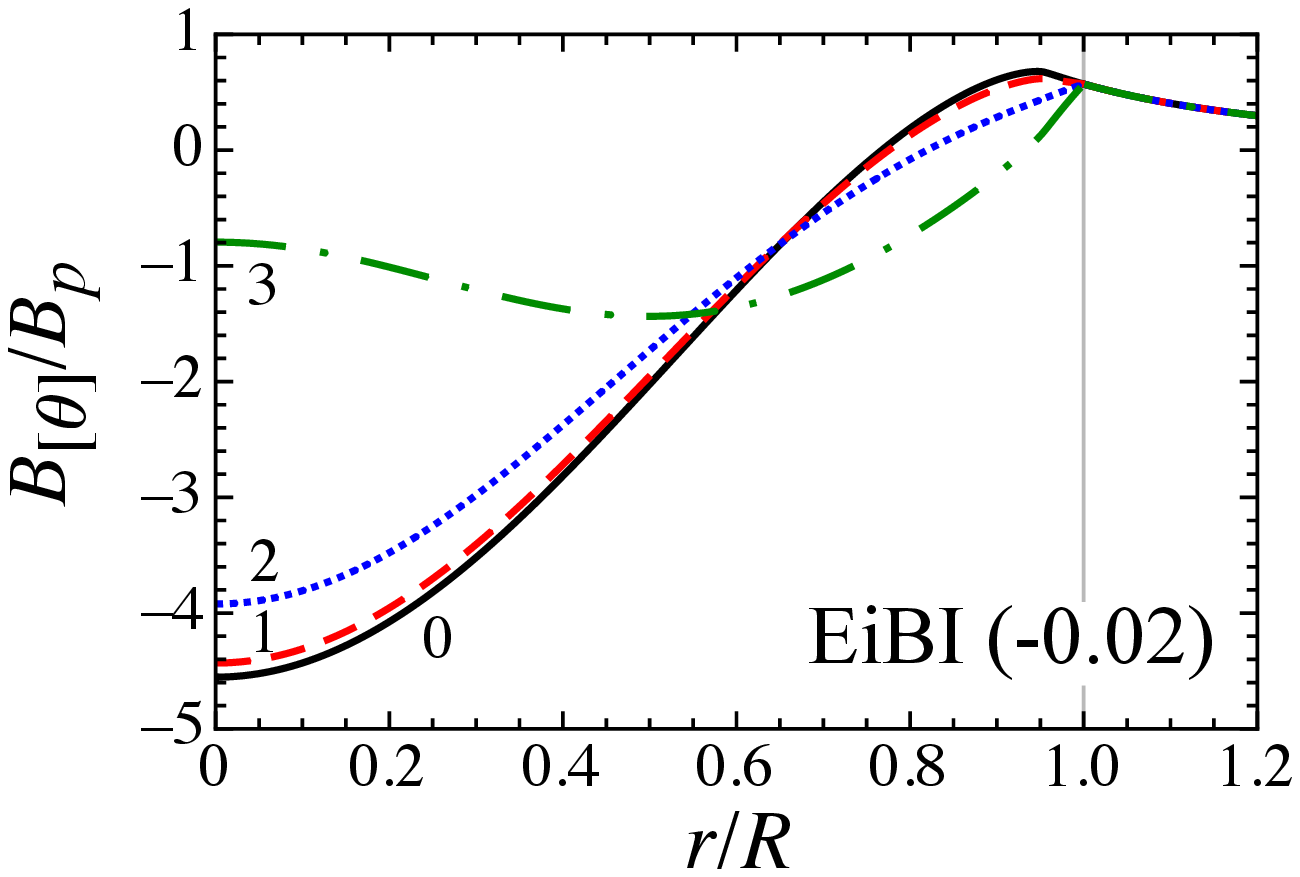} & \includegraphics[scale=0.42]{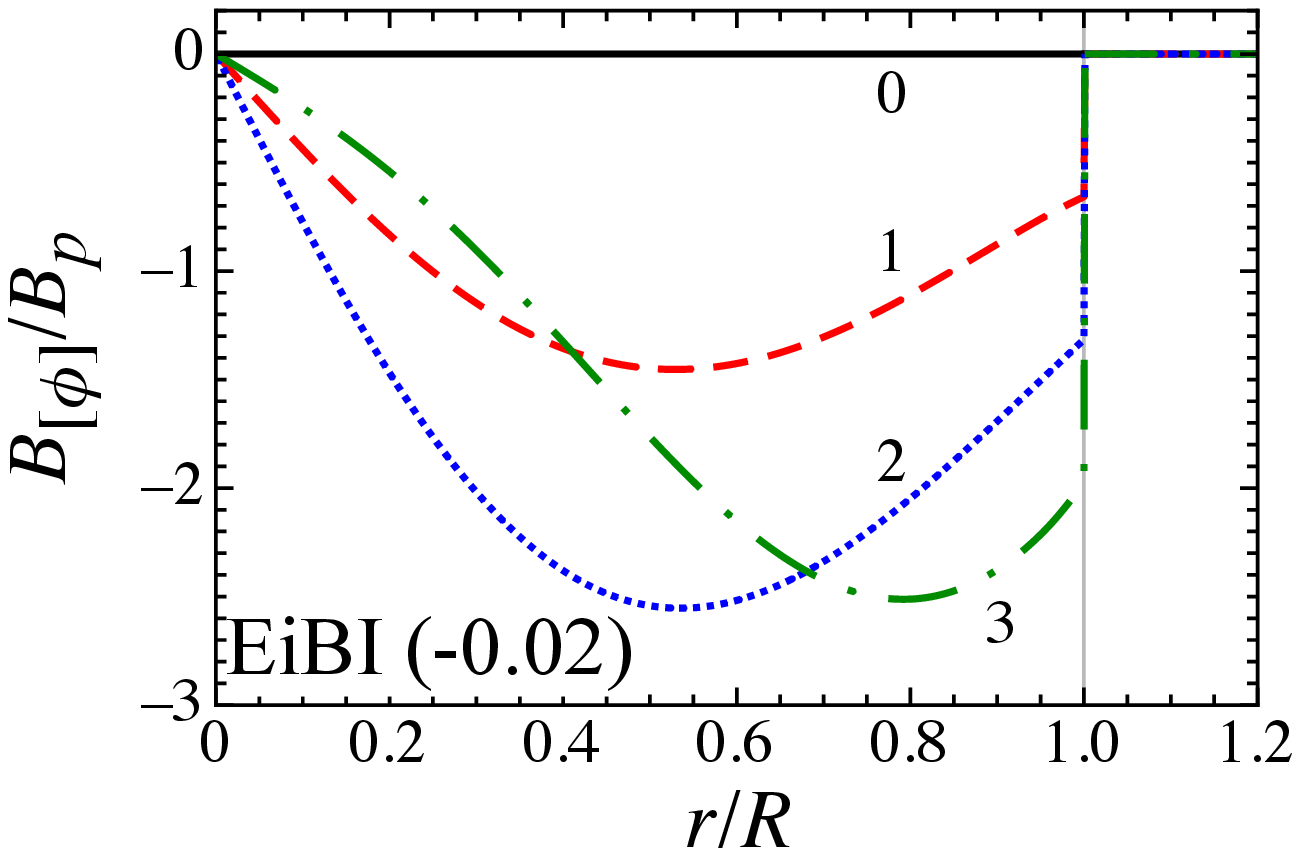} \\
\includegraphics[scale=0.42]{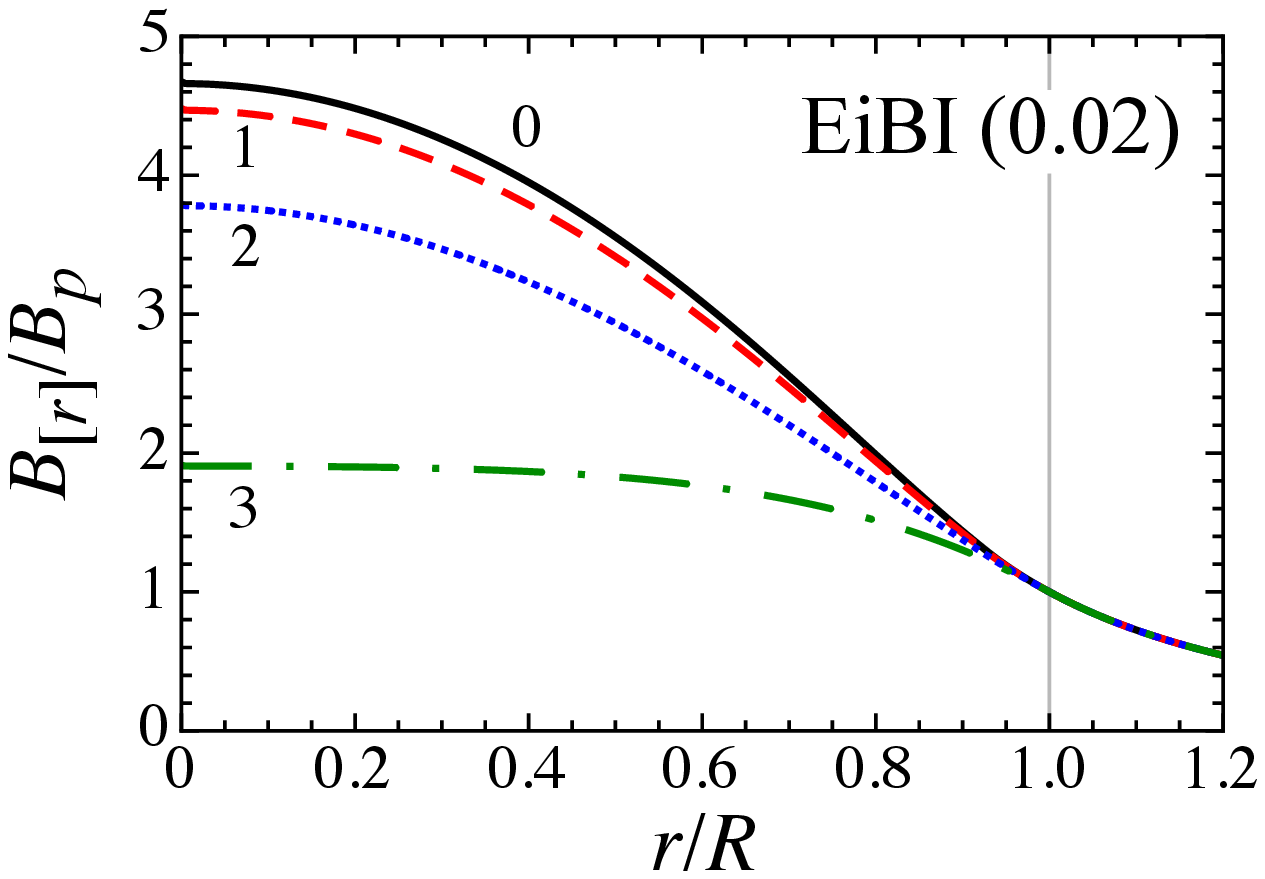} & \includegraphics[scale=0.42]{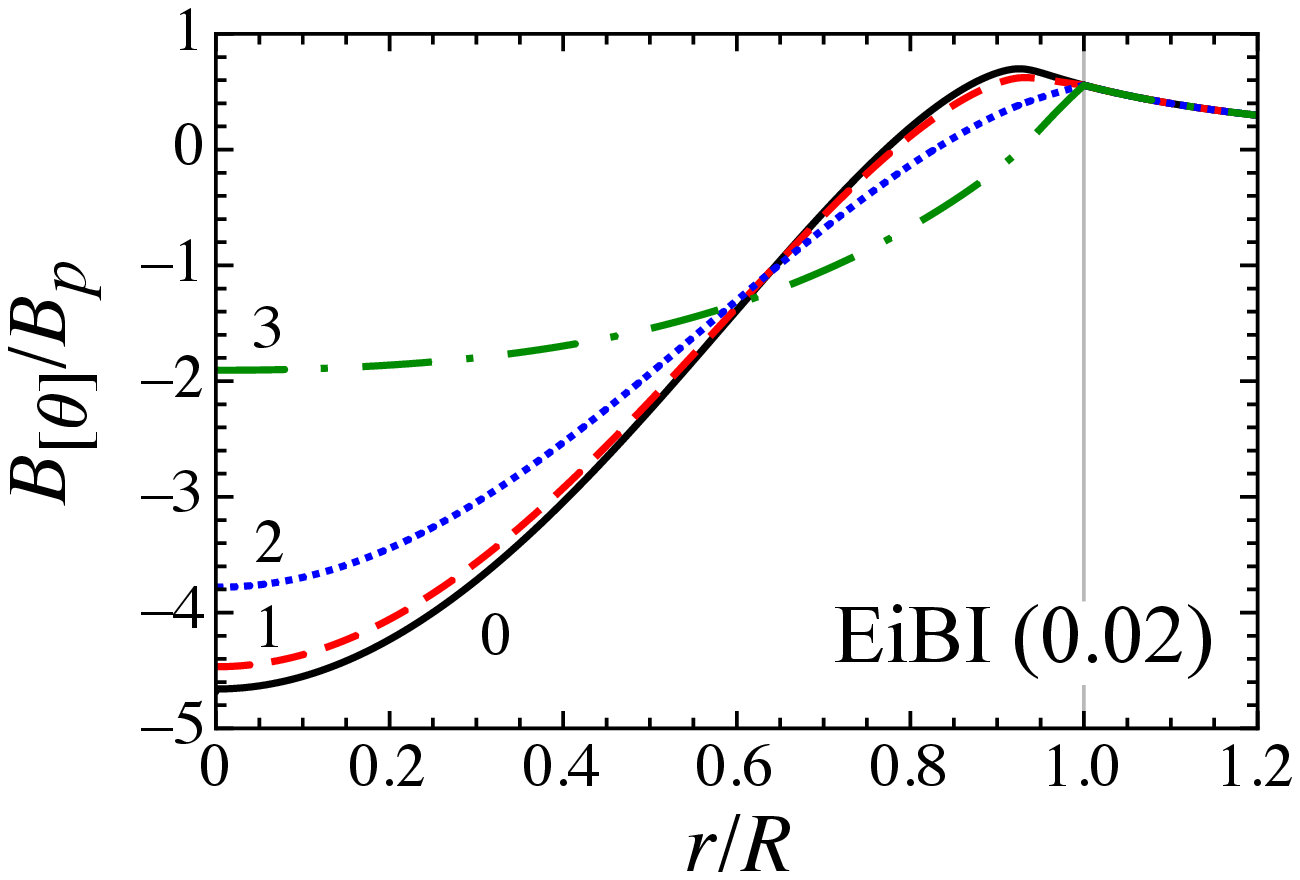} & \includegraphics[scale=0.42]{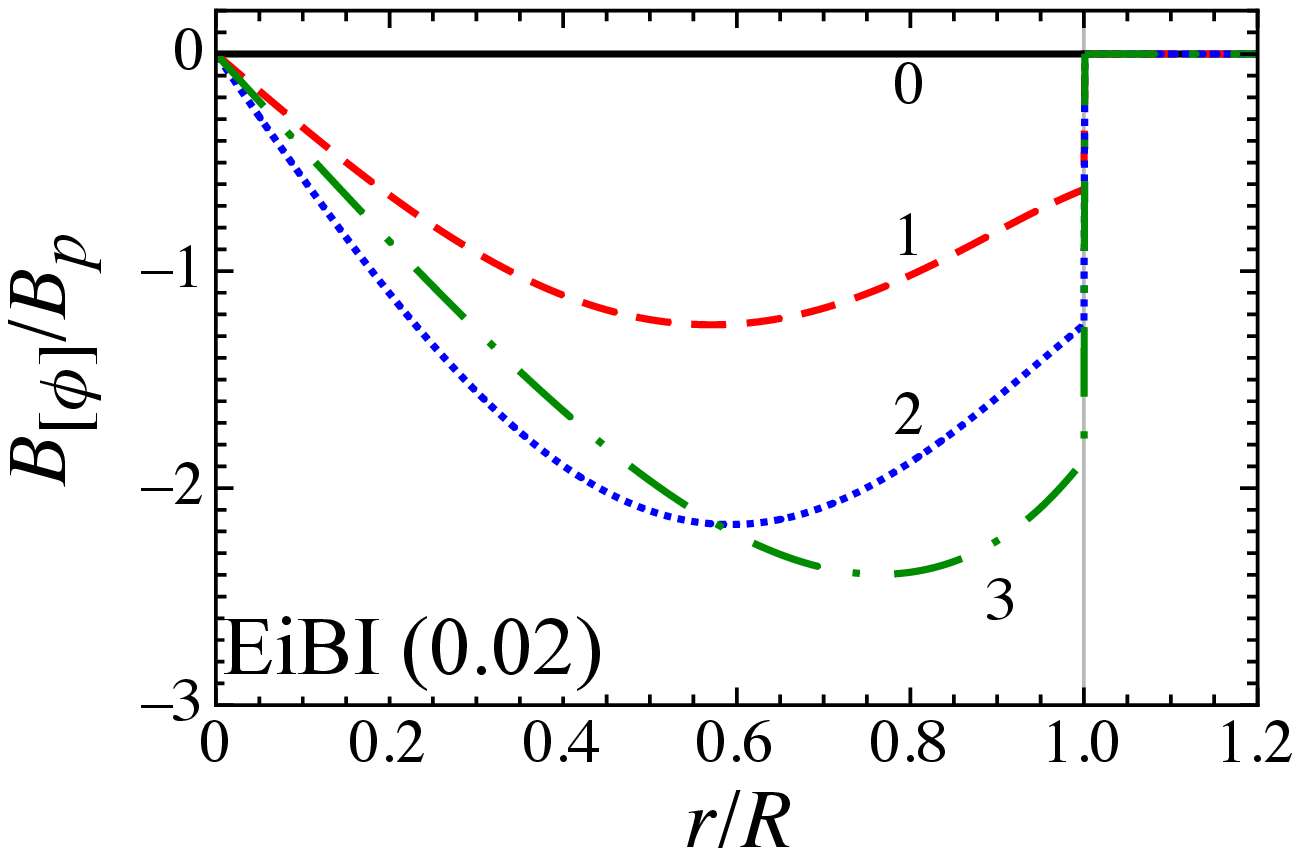} \\ 
\includegraphics[scale=0.42]{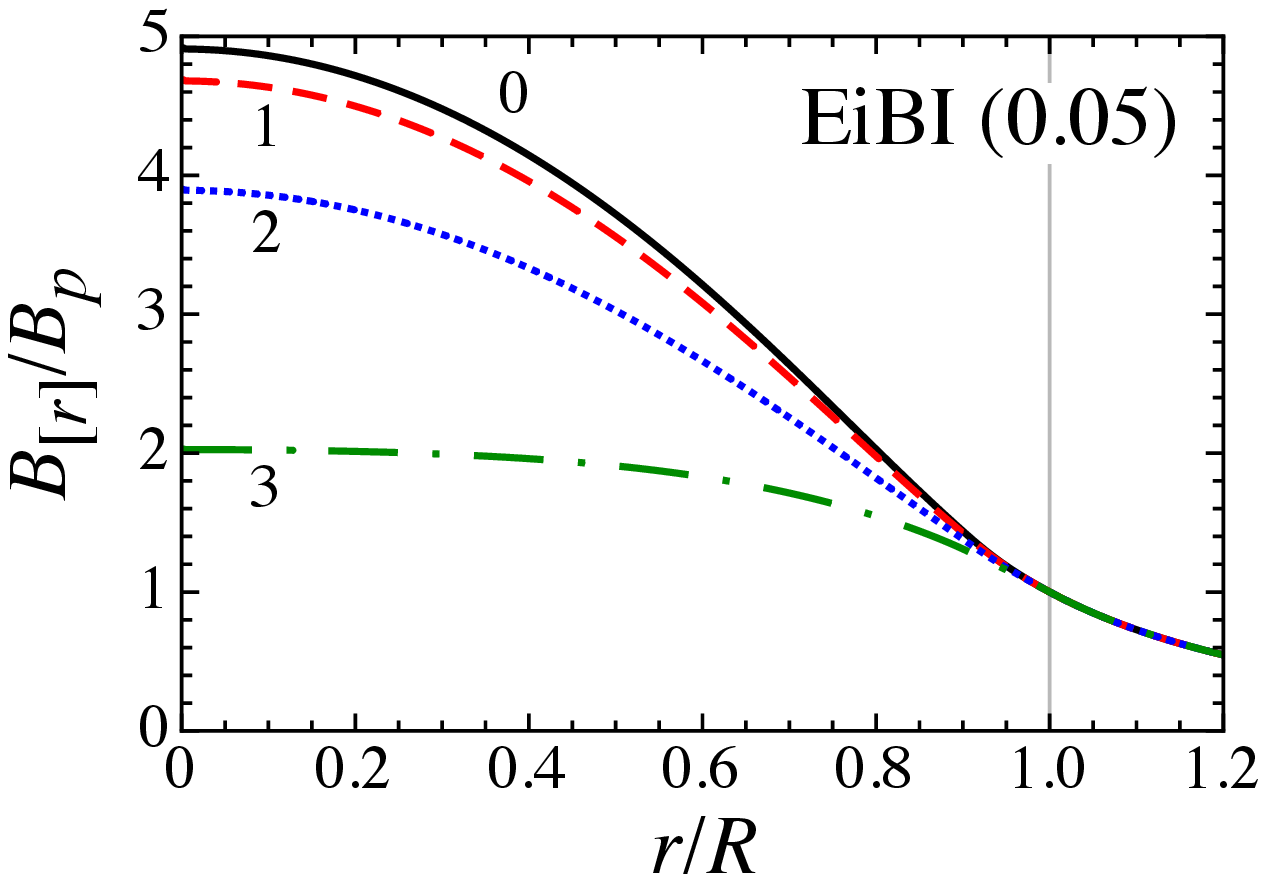} & \includegraphics[scale=0.42]{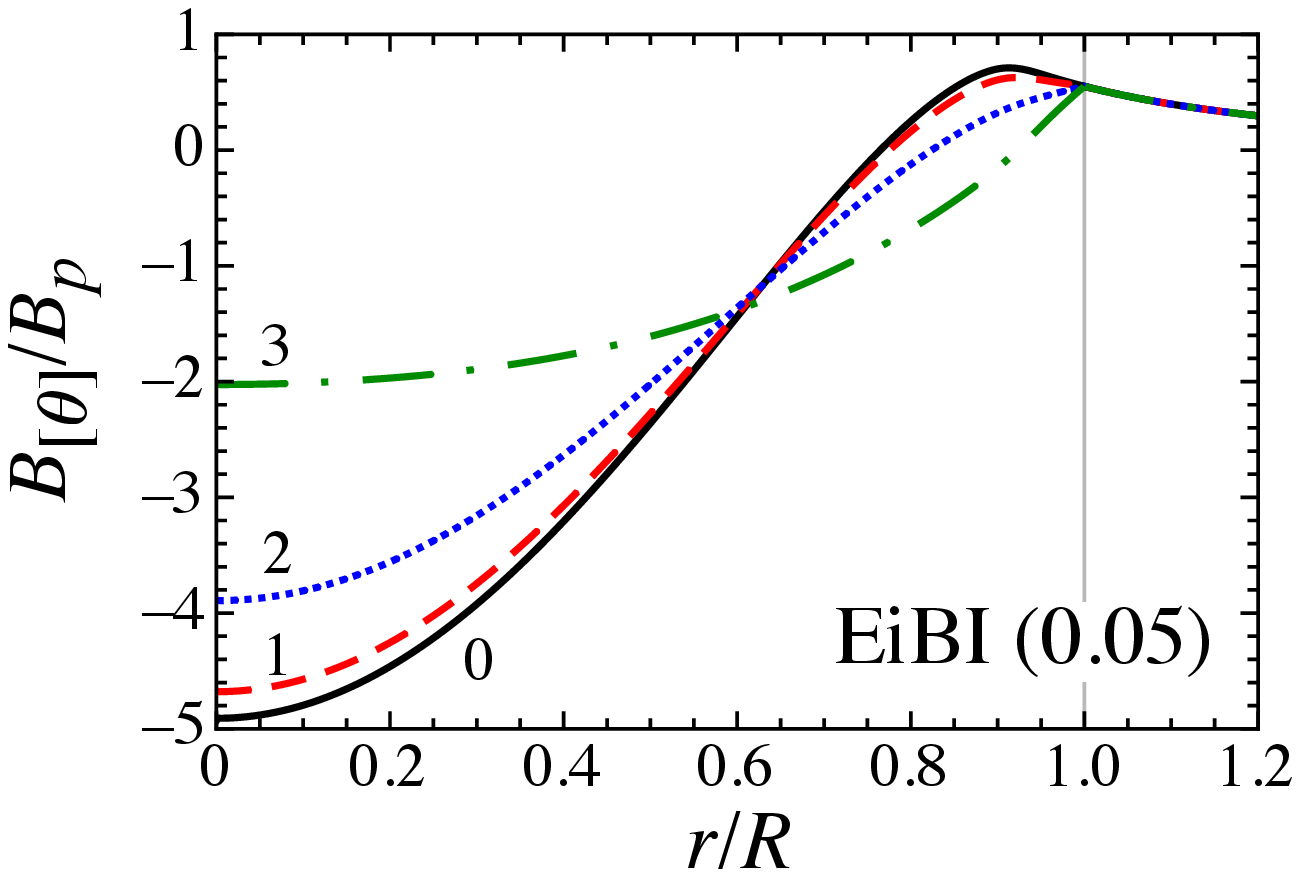} & \includegraphics[scale=0.42]{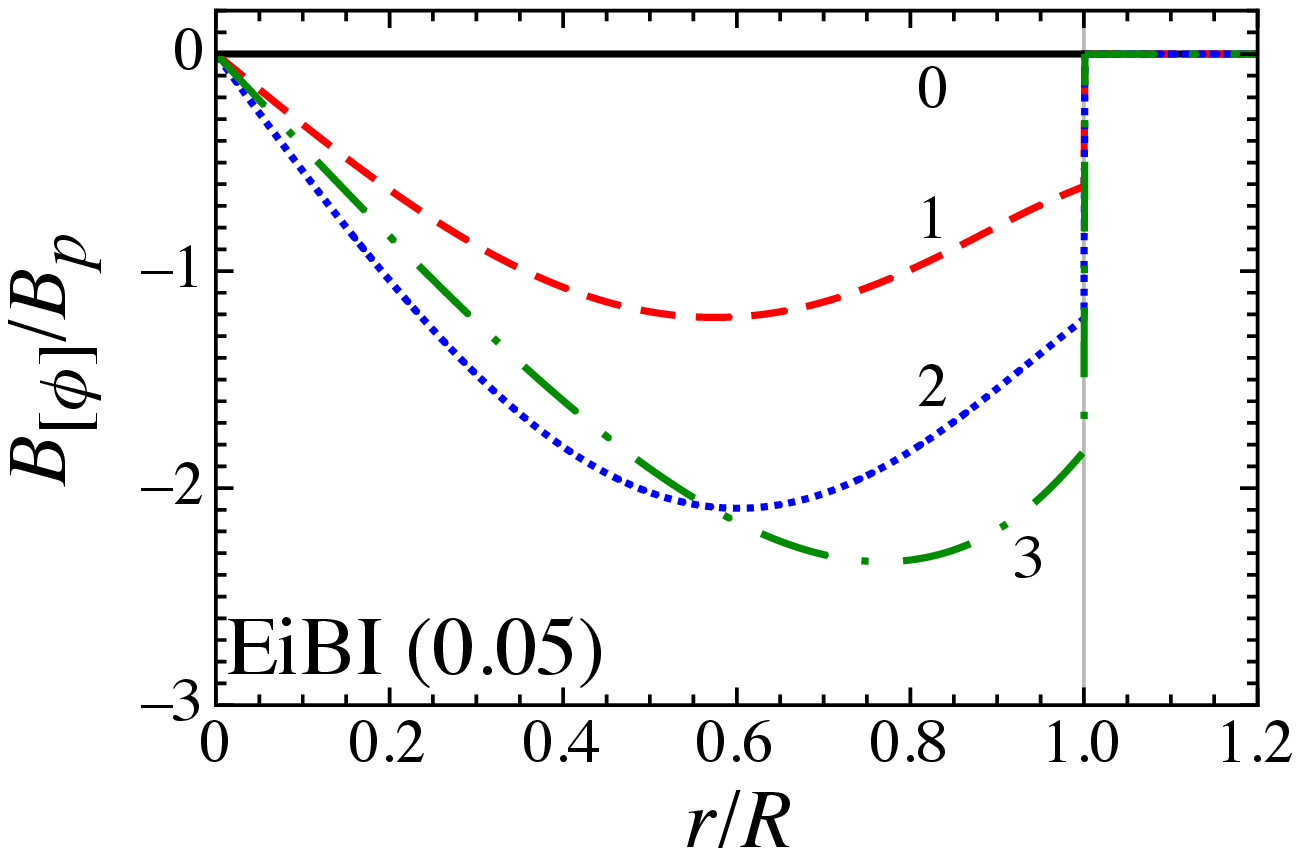} 
\end{tabular}
\end{center}
\caption{%%
Similar to Fig. \ref{fig:BM14T0}, but in EiBI with various coupling constants. Upper, middle, and lower panels correspond to the results in EiBI with $8\pi\kappa\varepsilon_s = -0.02$, 0.02, and 0.05, respectively.
}
\label{fig:BM14T00}
\end{figure*}

On the other hand, the magnetic distributions of $B_{[r]}/B_p$, $B_{[\theta]}/B_p$, and $B_{[\phi]}/B_p$ for the stellar models in EiBI with various coupling constants are shown in Fig. \ref{fig:BM14T00}, where the upper, middle, and lower panels correspond to the results with $8\pi\kappa\varepsilon_s=-0.02$, $0.02$, and $0.05$, respectively. Comparing this figure with Fig. \ref{fig:BM14T0}, one can see that the profiles of magnetic distributions in EiBI are basically similar to that in general relativity, where the distributions of $B_{[r]}/B_p$ and $B_{[\theta]}/B_p$ depend strongly on that of $B_{[\phi]}/B_p$. We also find that, as the coupling constant becomes smaller, the magnetic distributions are more sensitive to the value of $\zeta R$. For example, for $\zeta R=3$, one can see that $B_{[r]}/B_p$ and $|B_{[\theta]}/B_p|$ with $8\pi\kappa\varepsilon_s=-0.02$ in the vicinity of stellar center become smaller than that in general relativity. As a result, it is expected that the value of $\zeta_{\rm max}R$ for the stellar model with smaller coupling constant in EiBI could be smaller. In fact, we find that $\zeta_{\rm max}R=3.07$, $3.30$, $3.41$, and $3.51$ for the stellar models with $M=1.4M_\odot$ constructed with FPS EOS with $8\pi\kappa\varepsilon_s=-0.02$, $0$, $0.02$, and $0.05$, respectively.

%%%%%%%%%%%%%%%%%%%%%%%%%%%%%%%%%%%
 %Figure 8
%%%%%%%%%%%%%%%%%%%%%%%%%%%%%%%%%%%
%%%%%%%%%%%%%%%%%%%%%%%%%%%%%%%%%%%

\begin{figure*}
\begin{center}
\begin{tabular}{cc}
\includegraphics[scale=0.5]{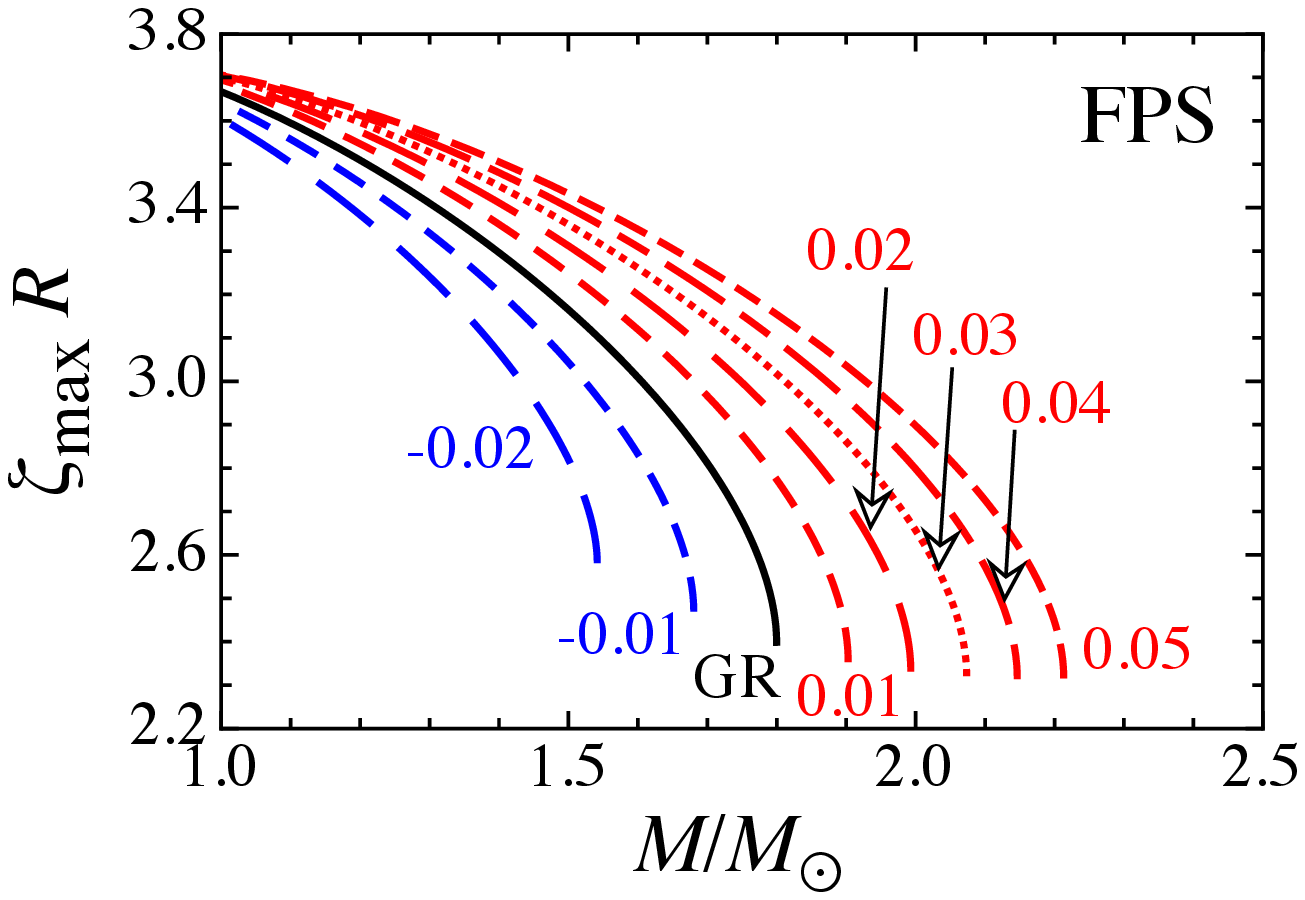} & 
\includegraphics[scale=0.5]{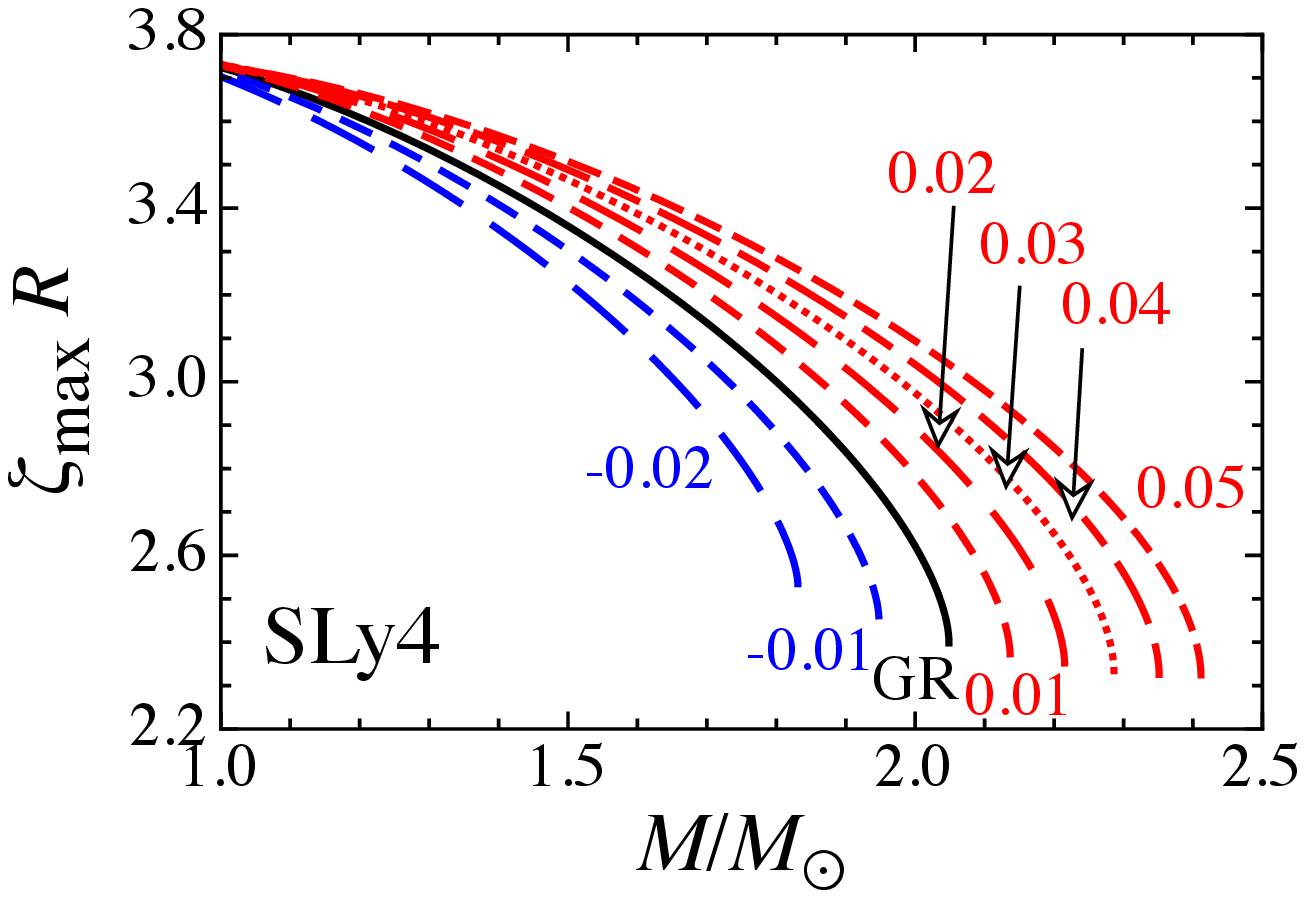}
\end{tabular}
\end{center}
\caption{%%
Maximum value of $\zeta$ allowed for the stellar models with various coupling constants in EiBI are shown as a function of the stellar mass, where the left and right panels correspond to the results for FPS and SLy4 EOSs, respectively. The labels in the figure denote the values of the coupling constant in EiBI.
}
\label{fig:zRM}
\end{figure*}

Furthermore, in Fig. \ref{fig:zRM}, we show the values of $\zeta_{\rm max}R$ for the stellar models with various stellar masses from $M=M_\odot$ up to the maximum mass, where the left and right panels correspond to the results for the stellar models constructed with FPS and SLy4 EOSs, respectively, and the labels in the figure denote the values of the coupling constant in EiBI. From this figure, one can see 
%that the normalized parameter $\zeta_{\max}R$ depends weakly on the EOS for neutron star matter. In particular, 
the value of $\zeta_{\max}R$ for low-mass neutron star model is almost independent of not only the EOS but also the coupling constant in EiBI. On the other hand, the magnetic fields for the stellar models with canonical mass are more or less dependent on both the EOS and the coupling constant in EiBI. That is, an uncertainty due to the EOS for neutron star matter is degenerate into that due to the coupling constant in EiBI. Thus, only the measurement of the magnetic properties for the neutron star with canonical mass might be insufficient to observationally distinguish  EiBI from general relativity. Anyway, the additional observations of the relativistic objects must become important to probe the gravitational theory in the strong field regime.

%%%%%%%%%%%%%%%%%%%%%%%%%%%%%%%%%%%
 %Figure 9
%%%%%%%%%%%%%%%%%%%%%%%%%%%%%%%%%%%
%%%%%%%%%%%%%%%%%%%%%%%%%%%%%%%%%%%

%%%%%%%%%%%%%%%%%%%%%%%%%%%%%%%%%%%
 %Figure 10
%%%%%%%%%%%%%%%%%%%%%%%%%%%%%%%%%%%
%\begin{figure*}
%\begin{center}
%\begin{tabular}{cc}
%\includegraphics[scale=0.5]{z-MF} & 
%\includegraphics[scale=0.5]{z-MS}
%\end{tabular}
%\end{center}
%\caption{%%
%$\zeta_{\rm max}$ for the stellar models with various coupling constants in EiBI are shown as a function of the stellar mass, where the left and right panels %correspond to the results for FPS and SLy4 EOSs, respectively. The meaning of the lines are the same as in Fig. \ref{fig:zRM}.
%}
%\label{fig:zM}
%\end{figure*}
%%%%%%%%%%%%%%%%%%%%%%%%%%%%%%%%%%%

%%%%%%%%%%%%%%%%%%%%%%%%%%%%%%%%%%%%%%%%%%%%%%%%
\section{Conclusion}
\label{sec:IV}
%%%%%%%%%%%%%%%%%%%%%%%%%%%%%%%%%%%%%%%%%%%%%%%%

We consider the magnetic fields in the neutron stars in EiBI, where we especially focus on the dipole magnetic fields because such fields must be dominant in old neutron stars. To construct magnetic fields inside the neutron star, we derive the relativistic Grad-Shafranov equation in EiBI. Since the spacetime in vacuum in EiBI is equivalent to that in general relativity, i.e., the Schwarzschild spacetime, the magnetic field in EiBI outside the star is also equivalent to that in general relativity. In such a way that the interior magnetic fields should be connected to the exterior solution, the structure of magnetic fields is determined. Then, we find that the magnetic geometry inside the neutron stars in EiBI is qualitatively similar to that in general relativity. The deviation of magnetic fields in EiBI from that in general relativity is not so much, which is almost comparable to the uncertainty due to the EOS for neutron star matter. Therefore, it might be difficult to distinguish EiBI from general relativity only by using the observations of the magnetic properties in neutron stars. However, the magnetic fields in the crust region for the neutron star with canonical mass depend weakly on the coupling constant in EiBI, while the crust properties such as the crust thickness depends strongly on the EOS for neutron star matter \cite{SKS2007}. That is, independently of the gravitational theory, one might be able to see the information about the EOS in crust region through the observations associated with the phenomena in crust region, such as the stellar oscillations. 
%On the other hand, we also find that some magnetic properties depend weakly on the EOS for neutron star matter, if the stellar mass of neutron star would be very low. Thus, it might be possible to discuss the gravitational theory via the observations of the low-mass neutron stars. 
Anyway, there are many uncertainties in the magnetic properties even in general relativity, such as the magnetic geometry and the current distribution supporting the fields, although we consider only dipole magnetic fields in this paper. Comparing to such uncertainties, the imprint of EiBI gravity on the magnetic fields is weak, which suggests that the magnetic field could be a poor probe of gravitational theories.

%\newpage
%%%%%%%%%%%%%%%%%%%%%%%%%%%%%%%%%%%%%%%%%%%%%%%%
\acknowledgments
%%%%%%%%%%%%%%%%%%%%%%%%%%%%%%%%%%%%%%%%%%%%%%%%
%We are grateful to E. Berti for his fruitful comments. 
This work was supported in part %by Grants-in-Aid for Scientific Research on Innovative Areas through No.\ 24105001 and No.\ 24105008 provided by MEXT, 
by Grant-in-Aid for Young Scientists (B) through No.\ 26800133 provided by JSPS. %, by the Yukawa International Program for Quark-hadron Sciences, and by the Grant-in-Aid for the global COE program ``The Next Generation of Physics, Spun from Universality and Emergence" from MEXT.

\appendix
%%%%%%%%%%%%%%%%%%%%%%%%%%%%%%%%%%%%%%%%%%%%%%%%
\section{Derivation of Eq. (\ref{eq:a1}) }   % Appendix A
\label{sec:appendix_1}
%%%%%%%%%%%%%%%%%%%%%%%%%%%%%%%%%%%%%%%%%%%%%%%%

According to Refs. \cite{CFGP2008,SCK2008}, we briefly show in this appendix how to derive Eq. (\ref{eq:a1}), which is the equation to determine the distribution of magnetic field inside the star. One can obtain the following equations from Eq. (\ref{eq:Max2})
\begin{gather}
  4\pi J^r = -\frac{1}{f}e^{-\lambda}\left(A_{r,\theta\theta} + A_{r,\theta}\frac{\cos\theta}{\sin\theta}\right), \label{eq:a3} \\
  4\pi J^\theta = \frac{1}{f}e^{-\lambda}\left[A_{r,\theta r} + A_{r,\theta}\left(\frac{\nu'}{2} - \frac{\lambda'}{2}\right)\right], \label{eq:a4}  \\
  4\pi J^\phi \sin^2\theta = -\frac{1}{f}e^{-\lambda}\left[A_{\phi,rr} + A_{\phi,r}\left(\frac{\nu'}{2} - \frac{\lambda'}{2}\right) \right]
     - \frac{1}{f^2}A_{\phi,\theta\theta} + \frac{1}{f^2}\frac{\cos\theta}{\sin\theta}A_{\phi,\theta},  \label{eq:a5}
\end{gather}
while from Eq. (\ref{eq:motion})
\begin{gather}
  -A_{r,\theta} J^\theta + A_{\phi,r} J^\phi = (\varepsilon + p)\frac{\nu'}{2} + p', \label{eq:a6} \\
  A_{r,\theta} J^r + A_{\phi,\theta}J^\phi = 0, \label{eq:a7} \\
  A_{\phi,r}J^r + A_{\phi,\theta}J^\theta = 0.  \label{eq:a8}
\end{gather}
Equation (\ref{eq:a8}) with Eqs. (\ref{eq:a3}) and (\ref{eq:a4}) can be written as
\begin{equation}
  -\eta_{,\theta} A_{\phi,r} + \eta_{,r} A_{\phi,\theta} = 0,
\end{equation}
where $\eta \equiv e^{\nu/2-\lambda/2} A_{r,\theta}\sin\theta$. Thus, $\eta$ should depend only on $A_\phi$ as $\eta=\zeta A_\phi$ with a constant $\zeta$. As a result, $A_r$ is expressed as $A_r = \zeta e^{-\nu/2+\lambda/2}a_\ell P_\ell$, if $A_\phi$ is expanded as Eq. (\ref{eq:expand}).

On the other hand, Eqs. (\ref{eq:a6}) and (\ref{eq:a7}) become
\begin{gather}
  \chi_{,r} = A_{\phi,r} {\cal J}, \label{eq:chir} \\
  \chi_{,\theta} = A_{\phi,\theta} {\cal J}, \label{eq:chitheta}
\end{gather}
where $\chi$ and ${\cal J}$ are defined as
\begin{gather}
  \chi_{,r} = (\varepsilon + p)\frac{\nu'}{2} + p', \\
  {\cal J} = \frac{1}{f(\varepsilon + p)\sin^2\theta}\left(J_\phi - \frac{\zeta^2}{4\pi}e^{-\nu}A_\phi\right). \
\end{gather}
Owing to the relation $\chi_{,r\theta} = \chi_{,\theta r}$ with Eqs. (\ref{eq:chir}) and (\ref{eq:chitheta}), one can obtain
\begin{equation}
  A_{\phi,r}{\cal J}_{,\theta} - A_{\phi,\theta}{\cal J}_{,r} = 0.
\end{equation}
Therefore, ${\cal J}$ also depends only on $A_\phi$, i.e., ${\cal J}=-c_0 - c_1 A_\phi$, where $c_0$ and $c_1$ are constants. That is,
\begin{equation}
  J_\phi = \frac{\zeta^2}{4\pi}e^{-\nu}A_\phi - (c_0 + c_1 A_\phi) f(\varepsilon + p)\sin^2\theta.  \label{eq:J}
\end{equation}
At last, substituting Eqs. (\ref{eq:expand}) and (\ref{eq:J}) into Eq. (\ref{eq:a5}), one can obtain the equation (\ref{eq:a1}) describing the function of $a_1$
%\begin{equation}
%  a_1'' + \left(\frac{\nu'}{2}  - \frac{\lambda'}{2}\right)a_1' + \left(\zeta^2 e^{-\nu} - \frac{2}{f}\right)e^{\lambda}a_1
%     = -4\pi e^{\lambda} j_1, 
%\end{equation}
%where $j_1=c_0 f(\varepsilon+p)$. 
We remark that the term of $c_1$ in Eq. (\ref{eq:J}) is neglected, because we focus on the dipole ($\ell=1$) magnetic distribution in this paper and the term of $c_1$ can contribute as the multipole higher than $\ell=3$ \cite{CFGP2008,SCK2008}.

%%%%%%%%%%%%%%%%%%%%%%%%%%%%%%%%%%%%%%%%%%%%%%%%

\end{document}